\documentclass[preprint,aps,pra,showpacs]{revtex4}

\usepackage{graphicx}
\usepackage{amsmath}
\usepackage{amsfonts}
\usepackage{amssymb}

\begin{document}

\author{Klaus Hornberger}
\author{John E.~Sipe}
\altaffiliation[Permanent address: ]{Department of Physics,
University of Toronto, 60 St.~George Street, Toronto, ON, M5S 1A7, Canada}
\affiliation{Universit\"{a}t Wien, Institut f\"{u}r Experimentalphysik,
Boltzmanngasse 5, 1090 Wien, Austria}

\title{Collisional decoherence reexamined}
\date{Apil 17, 2003}%\date{\today}
%03.65.-w   Quantum mechanics
%03.65.Ud   Entanglement and quantum nonlocality (e.g. EPR paradox,
%           Bell's inequalities, GHZ states, etc.)
%03.65.Ta   Foundations of quantum mechanics; measurement theory
%03.65.Yz   Decoherence; open systems; quantum statistical methods
%03.75.-b   Matter waves
%03.75.Dg   Atom and neutron interferometry

\begin{abstract}
We re-derive the quantum master equation for the decoherence of a
massive Brownian particle due to collisions with the lighter
particles from a thermal environment. Our careful treatment
avoids the occurrence of squares of Dirac delta functions. It
leads to a decoherence rate which is smaller by a factor of $2\pi$
compared to previous findings. This result, which is in
agreement with recent experiments, is confirmed by both a
physical analysis of the problem and by a perturbative
calculation in the weak coupling limit.
\end{abstract}
\maketitle

\section{Introduction}
\label{sec:intro}

A classic result of decoherence theory is the rapid decay in the
off-diagonal matrix elements in the coordinate representation of
the density operator $\rho (\mathbf{R}_1,\mathbf{R}_2;t)$ of a
massive Brownian particle suffering collisions with the lighter
particles of a thermal bath. Early calculations by Joos and Zeh
\cite{Joos1985a} were improved by later authors, and the result
of Gallis and Fleming \cite{Gallis1990a} seems to be the most
widely quoted \cite{Giulini1996a}. They find, in the limit of an
infinitely massive Brownian particle, that
\begin{equation}
\frac{\partial \rho (\mathbf{R}_1,\mathbf{R}_2;t)}{\partial t}=
-F(\mathbf{R}_1-\mathbf{R}_2)\rho (\mathbf{R}_1,\mathbf{R}_2;t),
\label{drhodt}
\end{equation}
where
\begin{equation}
F(\mathbf{R})=\varepsilon \, n\int_{0}^{\infty }dq\nu (q)\frac{q}{m}\int \frac{d%
\mathbf{\hat{n}}_{1}d\mathbf{\hat{n}}_{2}}{4\pi }\left( 1-e^{iq(\mathbf{\hat{%
n}}_{1}-\mathbf{\hat{n}}_{2})\cdot \mathbf{R}/\hbar }\right) \left| f(q%
\mathbf{\hat{n}}_{2},q\mathbf{\hat{n}}_{1})\right| ^{2},  \label{gamma}
\end{equation}
with $m$ the mass of the bath particles, $n$ their number density, and $%
\nu (q)\,dq$ the fraction of particles with momentum magnitude between $q$ and $q+dq$;
$\mathbf{\hat{n}}_{1}$ and $\mathbf{\hat{n}}_{2}$ are unit vectors,
with $d\mathbf{\hat{n}}_{1}$ and $d\mathbf{\hat{n}}_{2}$ the elements of
solid angle associated with them. The quantity $f(\mathbf{q}_{2},\mathbf{q}%
_{1})$ is the scattering amplitude of a bath particle off the Brownian
particle from initial momentum $\mathbf{q}_{1}$ to final momentum $\mathbf{q}%
_{2}$. Gallis and Fleming  find $\varepsilon =2\pi $.

We show here that this result is incorrect; the correct result is given by (%
\ref{gamma}) with $\varepsilon =1$. Needless to say, this
difference does not affect the qualitative conclusion that
off-diagonal elements decay exceedingly quickly for even
macroscopically small $| \mathbf{R}_1- \mathbf{R}_2|$.
Nonetheless, experimental techniques are now available that permit the
study of the quantum mechanical loss of coherence by collisions \cite{Hornberger2003a}.
Therefore not only a qualitative understanding of decoherence effects is
needed, but a quantitatively correct description is required as well.
Moreover, derivations of benchmark
equations in the theory of decoherence such
(\ref{drhodt},\ref{gamma}) illustrate the nature of the physics
and the assumptions involved, and uncovering the errors of
earlier results serve as cautionary tales that may facilitate  the
analysis of situations where the physics is more complicated.

In this paper we present two detailed calculations of the fundamental result (\ref
{drhodt},\ref{gamma}). The first is a scattering theory calculation in the
spirit of the usual derivations, but one that avoids a pitfall of those
calculations by using localized and normalized states in the scattering
calculation. The second is a weak coupling calculation that follows the
spirit of master equation derivations undertaken in, \textit{e.g., }%
quantum optics. In the first calculation we find (\ref{gamma}) with $%
\varepsilon =1$. In the second we find (\ref{gamma}) with $\varepsilon =1$
and $f(q\mathbf{\hat{n}}_{2},q\mathbf{\hat{n}}_{1})$ replaced by $f_{\rm B}(q%
\mathbf{\hat{n}}_{2},q\mathbf{\hat{n}}_{1})$, the first Born approximation
to that scattering amplitude. This is precisely what would be expected,
since the second calculation requires the assumption of weak interaction; it
thus serves to confirm the $\varepsilon =1$ result of the first. Neither of
these is the most elegant or general calculation one could imagine; the
first is rather cumbersome, and the second would be neater if generalized to
second quantized form \cite{Altenmuller1997a}.
But the first has the
advantage of displaying the physics of decoherence in an almost pictorial
way, while allowing a calculation involving the full scattering amplitude.
And the second, in its simple form, establishes a clear connection with the
usual approach to decoherence through the master equation approach common in
quantum optics. Totally separate in their approaches, we feel that together
they are a convincing demonstration that $\varepsilon =1$.

These two calculations are presented in sections \ref{sec:scatt}
and \ref{sec:weak} below. In section \ref{sec:dsquared} we return
to the traditional derivation and highlight its inherent
shortcomings. We show how it should be modified by using a simple
physical argument, which leads to a replacement rule for the
occurring square of a Dirac delta function. This  treatment then also
yields the result $\varepsilon =1$. Our concluding remarks are
presented in section \ref{sec:conclusions}.

\section{Scattering calculation}
\label{sec:scatt}

To set our notation we begin with a review of the standard
approach used to calculate collisional  decoherence. However, we
also wish to point out the difficulties that can arise in its
application, so we begin in a more detailed way than is normally
done.

To apply scattering theory in a careful way one has to begin with an
asymptotic-in state $\left| \phi _{m}\right\rangle \left| \psi \right\rangle
$, a normalized ket that is the direct product of a Brownian particle ket $%
\left| \phi _{m}\right\rangle $ and a bath particle ket $\left|
\psi \right\rangle $. The asymptotic-in ket is the result of the
evolution of a product ket $| \phi _{m}^{(-\infty
)}\rangle | \psi ^{(-\infty )}\rangle $ at
$t=-\infty $ to $t=0$ under the Hamiltonian that describes the
free evolution of both particles, without
interaction. The effect of the two-particle scattering
operator $\mathcal{S}$ on this asymptotic-in state,
$\mathcal{S}(| \phi _{m}\rangle| \psi \rangle )$, then produces
the asymptotic-out state. When evolved from $t=0$ to $t=\infty $
by the non-interacting Hamiltonian, the asymptotic-out state
yields the actual state at $t=\infty $ that evolves from $|
\phi _{m}^{(-\infty )}\rangle | \psi ^{(-\infty
)}\rangle $ at $t=-\infty $ under the influence of the full
Hamiltonian.

\begin{figure}
 \centering
  \includegraphics[width=0.75\linewidth]{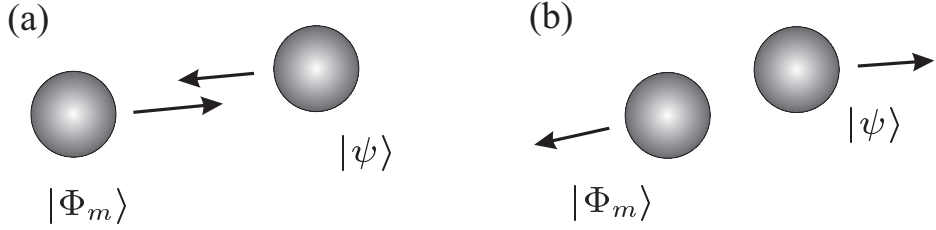}
  \caption{Sketched are the wave packets associated with $\left| \phi _{m}\right\rangle
$ and $\left| \psi \right\rangle $ at $t=0$. In configuration (a) the state $%
\left| \phi _{m}\right\rangle \left| \psi \right\rangle $ could be taken as
both an asymptotic-in state and an initial state at $t=0$; for configuration
(b) that would not be possible.}
  \label{fig:1}
\end{figure}

In general, of course, $\left| \phi _{m}\right\rangle \left| \psi
\right\rangle $ does not describe the \textit{actual} ket at $t=0$ that
evolves from $| \phi _{m}^{(-\infty )}\rangle | \psi
^{(-\infty )}\rangle $ at $t=-\infty $, because the evolution of that
actual ket involves the particle interaction. But if the kets $\left| \phi
_{m}\right\rangle $ and $\left| \psi \right\rangle $ are such that the
(short-range) interaction between the particles has not yet had an effect (%
\textit{e.g., }Fig. 1a but not Fig. 1b), then $\left| \phi
_{m}\right\rangle \left| \psi \right\rangle $ can be taken as the
\textit{actual} ket at $t=0$ as well as the asymptotic-in ket. We
only consider kets $\left| \phi _{m}\right\rangle $ and $\left|
\psi \right\rangle $ of this form below.

We now turn to the impending collision of a bath particle characterized by
$\left| \psi \right\rangle $ and a Brownian particle described by
a reduced density operator at $t=0$ given by a convex sum of projectors $|
\phi _{m}\rangle\langle\phi _{m}| $,
\begin{eqnarray*}
\rho _{\rm in} &=&\sum_{m}\mathsf{p}_{m}\left| \phi _{m}\right\rangle
\left\langle \phi _{m}\right|  \\
&=&\int d\mathbf{R}_{1}d\mathbf{R}_{2}\;\left| \mathbf{R}_{1}\right\rangle
\rho _{o}(\mathbf{R}_{1}\mathbf{,R}_{2})\left\langle \mathbf{R}_{2}\right| ,
\end{eqnarray*}
with probabilities $\mathsf{p}_{m} >0$, $\sum\mathsf{p}_{m} =1$.
Here the
$\left| \mathbf{R}_{1,2}\right\rangle $ label position eigenkets of
the Brownian particle, and
\begin{equation}
\rho _{o}(\mathbf{R}_{1}\mathbf{,R}_{2})=\sum_{m}\mathsf{p}_{m}\left\langle
\mathbf{R}_{1}\mathbf{|}\phi _{m}\right\rangle \left\langle \phi _{m}|%
\mathbf{R}_{2}\right\rangle   \label{rhonought}
\end{equation}
its position representation. Then
\begin{equation}
\rho _{\rm in}^{\rm total}=\rho _{\rm in}\otimes \left| \psi \right\rangle
\left\langle \psi \right|    \label{rhoin}
\end{equation}
can be considered both as the full initial (at $t=0$) density operator, and
the full asymptotic-in density operator. The full asymptotic-out density
operator is then
\begin{eqnarray*}
\rho _{\rm out}^{\rm total} &=&\mathcal{S}\rho _{\rm in}^{\rm total}\mathcal{S}^{\dagger } \\
&=&\int d\mathbf{R}_{1}d\mathbf{R}_{2}\;\mathcal{S}\left( \left| \mathbf{R}%
_{1}\right\rangle \left| \psi \right\rangle \right) \rho _{o}(\mathbf{R}_{1}%
\mathbf{,R}_{2})\left( \left\langle \psi \right| \left\langle \mathbf{R}%
_{2}\right| \right) \mathcal{S}^{\dagger }.
\end{eqnarray*}
To determine terms such as $\mathcal{S}\left( \left| \mathbf{R}\right\rangle
\left| \psi \right\rangle \right) $it is useful to first consider the effect
of the $\mathcal{S}$ operator on direct products $\left| \mathbf{P}%
\right\rangle \left| \mathbf{p}\right\rangle $ of eigenkets $\left| \mathbf{P%
}\right\rangle $ of the Brownian particle momentum and eigenkets $\left|
\mathbf{p}\right\rangle $ of the bath particle momentum. Since the total
momentum commutes with the $\mathcal{S}$ operator the scattering
transformation can be reduced to a one-particle problem, with
\[
\mathcal{S}\left( \left| \mathbf{P}\right\rangle \left| \mathbf{p}%
\right\rangle \right) =\int d\mathbf{q}\left| \mathbf{P}-\mathbf{q}%
\right\rangle \left| \mathbf{p+q}\right\rangle \left\langle \frac{m^*}{m}%
\mathbf{p-}\frac{m^*}{M}\mathbf{\mathbf{P}+q}|\mathcal{S}_{o}|\frac{%
m^*}{m}\mathbf{p-}\frac{m^*}{M}\mathbf{P}\right\rangle ,
\]
where the matrix element here is that of the \textit{one-}particle
scattering operator $\mathcal{S}_{o}$ corresponding to the two-body
interaction acting in the Hilbert space of the bath particle, and $%
m^*=mM/(m+M)$ is the reduced mass. In the limit that the Brownian
particle is much more massive than the bath particle, $M\gg m$, this reduces
to
\[
\mathcal{S}\left( \left| \mathbf{P}\right\rangle \left| \mathbf{p}%
\right\rangle \right) \rightarrow \int d\mathbf{q}\left| \mathbf{P}-\mathbf{q%
}\right\rangle \left| \mathbf{p+q}\right\rangle \left\langle \mathbf{p+q}|%
\mathcal{S}_{o}|\mathbf{p}\right\rangle
\]
or, moving to a position representation for the Brownian particle,
\begin{eqnarray*}
\mathcal{S}\left( \left| \mathbf{R}\right\rangle \left| \mathbf{p}%
\right\rangle \right)  &=&\int d\mathbf{q}\left| \mathbf{R}\right\rangle
e^{-i\mathbf{q\cdot R/\hbar }}\left| \mathbf{p+q}\right\rangle \left\langle
\mathbf{p+q}|\mathcal{S}_{o}|\mathbf{p}\right\rangle  \\
&=&\int d\mathbf{q}\left| \mathbf{R}\right\rangle \left| \mathbf{p+q}%
\right\rangle \left\langle \mathbf{p+q}|e^{-i\mathcal{\mathfrak{p}\cdot }\mathbf{%
R}/\hbar }\mathcal{S}_{o}e^{i\mathcal{\mathfrak{p}\cdot }\mathbf{R}/\hbar }|%
\mathbf{p}\right\rangle  \\
&=&\left| \mathbf{R}\right\rangle \left( e^{-i\mathcal{\mathfrak{p}\cdot }%
\mathbf{R}/\hbar }\mathcal{S}_{o}e^{i\mathcal{\mathfrak{p}\cdot }\mathbf{R}%
/\hbar }\left| \mathbf{p}\right\rangle \right) ,
\end{eqnarray*}
where $\mathfrak{p}$ is the momentum operator for the bath particle, and
so for general states $|\psi\rangle$
\begin{eqnarray*}
\mathcal{S}\left( \left| \mathbf{R}\right\rangle \left| \psi \right\rangle
\right)  &=&\left| \mathbf{R}\right\rangle \left( e^{-i\mathcal{\mathfrak{p}%
\cdot }\mathbf{R}/\hbar }\mathcal{S}_{o}e^{i\mathcal{\mathfrak{p}\cdot }\mathbf{R%
}/\hbar }\left| \psi \right\rangle \right)  \\
&\equiv &\left| \mathbf{R}\right\rangle \left| \psi ^{\mathbf{R}%
}\right\rangle ,
\end{eqnarray*}
where
\[
\left| \psi ^{\mathbf{R}}\right\rangle =e^{-i\mathcal{\mathfrak{p}\cdot }\mathbf{%
R}/\hbar }\mathcal{S}_{o}e^{i\mathcal{\mathfrak{p}\cdot }\mathbf{R}/\hbar
}\left| \psi \right\rangle ,
\]
and thus
\[
\rho _{\rm out}^{\rm total}=\int d\mathbf{R}_{1}d\mathbf{R}_{2}\;\left| \mathbf{R}%
_{1}\right\rangle \left| \psi ^{\mathbf{R}_{1}}\right\rangle \rho _{o}(%
\mathbf{R}_{1}\mathbf{,R}_{2})\left\langle \psi ^{\mathbf{R}_{2}}\right|
\left\langle \mathbf{R}_{2}\right| .
\]
Although $\rho _{\rm out}^{\rm total}$ is not the final density operator at $%
t=\infty $, but only the asymptotic-out density operator, it evolves to the
final density operator through the non-interacting Hamiltonian, and overlaps
of the form $\left\langle \psi ^{\mathbf{R}_{2}}|\psi ^{\mathbf{R}%
_{1}}\right\rangle $ will be preserved during this free evolution. So the
final reduced density operator for the Brownian particle at $t=\infty $ is
\begin{eqnarray*}
\rho _{\rm final} &=&\int d\mathbf{R}_{1}d\mathbf{R}_{2}\;\left| \mathbf{R}%
_{1}\right\rangle \left\langle \psi ^{\mathbf{R}_{2}}|\psi ^{\mathbf{R}%
_{1}}\right\rangle \rho _{o}(\mathbf{R}_{1}\mathbf{,R}_{2})\left\langle
\mathbf{R}_{2}\right|  \\
&\equiv &\int d\mathbf{R}_{1}d\mathbf{R}_{2}\;\left| \mathbf{R}%
_{1}\right\rangle \rho (\mathbf{R}_{1}\mathbf{,R}_{2})\left\langle \mathbf{R}%
_{2}\right| ,
\end{eqnarray*}
where
\begin{eqnarray}
\rho (\mathbf{R}_{1}\mathbf{,R}_{2})=\left\langle \psi ^{\mathbf{R}%
_{2}}|\psi ^{\mathbf{R}_{1}}\right\rangle \rho _{o}(\mathbf{R}_{1}\mathbf{,R}%
_{2}).
\label{eq:rhochange}
\end{eqnarray}
As is well understood, decoherence arises because the bath particle becomes
entangled with the Brownian particle and the two (asymptotic-out) states $\left|
\psi ^{\mathbf{R}_{2}}\right\rangle $ and $\left| \psi ^{\mathbf{R}%
_{1}}\right\rangle $ resulting from scattering interactions associated with
the same bath ket $\left| \psi \right\rangle $ and different position
eigenkets $\left| \mathbf{R}_{2}\right\rangle $ and $\left| \mathbf{R}%
_{1}\right\rangle $ can have negligible overlap even for $\left| \mathbf{R}%
_{2}\mathbf{-R}_{1}\right| $ small. The change of the Brownian particle's
reduced density operator by a single collision is
\begin{eqnarray}
\Delta \rho (\mathbf{R}_{1}\mathbf{,R}_{2}) &\equiv &\rho (\mathbf{R}_{1}%
\mathbf{,R}_{2})-\rho _{o}(\mathbf{R}_{1}\mathbf{,R}_{2}) \nonumber\\
&=&\left( \left\langle \psi ^{\mathbf{R}_{2}}|\psi ^{\mathbf{R}%
_{1}}\right\rangle -1\right) \rho _{o}(\mathbf{R}_{1}\mathbf{,R}_{2})
\;.
\end{eqnarray}
It involves overlap terms of the form
\begin{eqnarray}
\langle \psi ^{\mathbf{R}_{2}}|\psi ^{\mathbf{R}_{1}}\rangle
&=&\langle \psi |e^{-i\mathcal{\mathfrak{p}\cdot }\mathbf{R}_{2}/\hbar }%
\mathcal{S}_{o}^{\dagger }e^{-i\mathcal{\mathfrak{p}\cdot (}\mathbf{R}_{1}%
\mathbf{-R}_{2})/\hbar }\mathcal{S}_{o}e^{i\mathcal{\mathfrak{p}\cdot }\mathbf{R}%
_{1}/\hbar }|\psi \rangle  \nonumber\\
&=&\langle \psi |\mathcal{S}_{2}^{\dagger }\mathcal{S}_{1}|\psi
\rangle
=\mathrm{tr}_{\rm bath}\big\{\mathcal{S}_{2}^{\dagger
}\mathcal{S}_{1}|\psi\rangle\langle\psi|\big\}
\;,
\label{eq:overlap}
\end{eqnarray}
where the operators
\begin{equation}
\mathcal{S}_{j}=e^{-i\mathcal{\mathfrak{p}\cdot }\mathbf{R}_{j}/\hbar }\mathcal{S%
}_{o}e^{i\mathcal{\mathfrak{p}\cdot }\mathbf{R}_{j}/\hbar }
\label{Smatrices}
\end{equation}
for $j=1,2$ are translated scattering operators. We introduce corresponding
$\mathcal{T}_{j}$ operators according to
\begin{equation}
\mathcal{S}_{j}=1+i\mathcal{T}_{j},  \label{Tmatrices}
\end{equation}
and using the unitarity of the $\mathcal{S}_{j}$, which follows immediately
because $\mathcal{S}_{o}$ is unitary, we find
\begin{eqnarray*}
\mathcal{S}_{2}^{\dagger }\mathcal{S}_{1} &=&1+\mathcal{T}_{2}^{\dagger }%
\mathcal{T}_{1}-\frac{1}{2}\mathcal{T}_{1}^{\dagger }\mathcal{T}_{1}-\frac{1%
}{2}\mathcal{T}_{2}^{\dagger }\mathcal{T}_{2} \\
&&+\frac{i}{2}\left( \mathcal{T}_{1}+\mathcal{T}_{1}^{\dagger }\right) -%
\frac{i}{2}\left( \mathcal{T}_{2}+\mathcal{T}_{2}^{\dagger }\right)
\end{eqnarray*}
and so
\begin{equation}
\left\langle \psi ^{\mathbf{R}_{2}}|\psi ^{\mathbf{R}_{1}}\right\rangle
=1+\left\langle \psi |\mathcal{A}|\psi \right\rangle ,  \label{overlapterm}
\end{equation}
where
\begin{eqnarray*}
\mathcal{A} &=&\mathcal{T}_{2}^{\dagger }\mathcal{T}_{1}-\frac{1}{2}\mathcal{%
T}_{1}^{\dagger }\mathcal{T}_{1}-\frac{1}{2}\mathcal{T}_{2}^{\dagger }%
\mathcal{T}_{2} \\
&&+\frac{i}{2}\left( \mathcal{T}_{1}+\mathcal{T}_{1}^{\dagger }\right) -%
\frac{i}{2}\left( \mathcal{T}_{2}+\mathcal{T}_{2}^{\dagger }\right) .
\end{eqnarray*}
Thus the change in the Brownian particle reduced density operator is
\begin{equation}
\Delta \rho (\mathbf{R}_{1}\mathbf{,R}_{2})=\left\langle \psi |\mathcal{A}%
|\psi \right\rangle \rho _{o}(\mathbf{R}_{1}\mathbf{,R}_{2})
\label{deltarho}
\end{equation}
The general strategy is to evaluate the matrix element  $%
\left\langle \psi |\mathcal{A}|\psi \right\rangle $ by inserting complete
sets of momentum eigenstates,
\begin{equation}
\left\langle \psi |\mathcal{A}|\psi \right\rangle =\int d\mathbf{q}_{1}d%
\mathbf{q}_{2}\left\langle \psi |\mathbf{q}_{2}\right\rangle \left\langle
\mathbf{q}_{2}|\mathcal{A}|\mathbf{q}_{1}\right\rangle \left\langle \mathbf{q%
}_{1}|\psi \right\rangle
\;,
\label{momentumdecomp}
\end{equation}
determine $\left\langle \mathbf{q}_{2}|\mathcal{A}|\mathbf{q}%
_{1}\right\rangle $, and then perform the momentum eigenstate
integrals. Writing $\mathcal{S}_{o}=1+i\mathcal{T}_{o}$ as well,
and using the relations (\ref{Smatrices}) and (\ref{Tmatrices})
we find
\begin{eqnarray}
\left\langle \mathbf{q}_{2}|\mathcal{A}|\mathbf{q}_{1}\right\rangle  &=&e^{i(%
\mathbf{q}_{1}\cdot \mathbf{R}_{1}-\mathbf{q}_{2}\cdot \mathbf{R}_{2})/\hbar
}\left\langle \mathbf{q}_{2}|\mathcal{T}_{o}^{\dagger }e^{i\mathfrak{p}\cdot (%
\mathbf{R}_{2}-\mathbf{R}_{1})/\hbar }\mathcal{T}_{o}|\mathbf{q}%
_{1}\right\rangle   \label{A21} \\
&&-\frac{1}{2}e^{i(\mathbf{q}_{1}-\mathbf{q}_{2})\cdot \mathbf{R}_{1}/\hbar
}\left\langle \mathbf{q}_{2}|\mathcal{T}_{o}^{\dagger }\mathcal{T}_{o}|%
\mathbf{q}_{1}\right\rangle   \nonumber \\
&&-\frac{1}{2}e^{i(\mathbf{q}_{1}-\mathbf{q}_{2})\cdot \mathbf{R}_{2}/\hbar
}\left\langle \mathbf{q}_{2}|\mathcal{T}_{o}^{\dagger }\mathcal{T}_{o}|%
\mathbf{q}_{1}\right\rangle   \nonumber \\
&&+\frac{i}{2}\left[ e^{i(\mathbf{q}_{1}-\mathbf{q}_{2})\cdot \mathbf{R}%
_{1}/\hbar }-e^{i(\mathbf{q}_{1}-\mathbf{q}_{2})\cdot \mathbf{R}_{2}/\hbar
}\right] \left\langle \mathbf{q}_{2}|\mathcal{T}_{o}+\mathcal{T}%
_{o}^{\dagger }|\mathbf{q}_{1}\right\rangle .  \nonumber
\end{eqnarray}
Since the $\mathcal{S}_{o}$ operator matrix elements are given by
\cite{Taylor1972a}
\begin{equation}
\left\langle \mathbf{q}_{2}|\mathcal{S}_{o}|\mathbf{q}_{1}\right\rangle
=\delta (\mathbf{q}_{2}-\mathbf{q}_{1})+\frac{i}{2\pi \hbar m}\delta
(E_{2}-E_{1})f(\mathbf{q}_{2},\mathbf{q}_{1}),  \label{Snought}
\end{equation}
where $f(\mathbf{q}_{2},\mathbf{q}_{1})$ is the scattering amplitude, we can
identify
\begin{eqnarray}
\left\langle \mathbf{q}_{2}|\mathcal{T}_{o}|\mathbf{q}_{1}\right\rangle  &=&%
\frac{1}{2\pi \hbar m}\delta (E_{2}-E_{1})f(\mathbf{q}_{2},\mathbf{q}_{1})
\label{Tnought} \\
&=&\frac{\delta (q_{2}-q_{1})}{2\pi \hbar q_{2}}f(\mathbf{q}_{2},\mathbf{q}%
_{1}),  \nonumber
\end{eqnarray}
where $E_{i}=q_{i}^{2}/(2m).$

Now in the traditional calculations \cite{Joos1985a,Gallis1990a,Giulini1996a}
one calculates $\partial \rho (\mathbf{R}%
_{1},\mathbf{R}_{2})/\partial t$ by considering the change $\Delta \rho (\mathbf{R}_{1}%
\mathbf{,R}_{2})$ in a time $\Delta t$ due to collisions with bath particles
that would pass in the neighborhood of the Brownian particle, taking the
distribution of their velocities from the assumed thermal equilibrium of the
bath. To calculate $\Delta \rho (\mathbf{R}_{1}\mathbf{,R}_{2})$ from one of
these bath particles, a box-normalized momentum eigenstate, $\widetilde{|\mathbf{q}\rangle}$
  is used in place of a localized ket $\left| \psi \right\rangle
$. Unlike the $\left| \phi _{m}\right\rangle \left| \psi
\right\rangle $ states we introduced above, the $\left| \phi
_{m}\right\rangle \widetilde{|\mathbf{q}\rangle} $ obviously
cannot be considered either as asymptotic-in states or as the
actual states at $t=0$ since the $\widetilde{|\mathbf{q}\rangle}$
are delocalized. Nonetheless,  the traditional approach seems to
simplify the calculation
because, as is clear from (\ref{momentumdecomp}), only diagonal elements $%
\left\langle \mathbf{q}|\mathcal{A}|\mathbf{q}\right\rangle $ are required
if the limit of an infinite box is taken. But from the expression (\ref
{A21}) for $\left\langle \mathbf{q}_{2}|\mathcal{A}|\mathbf{q}%
_{1}\right\rangle $ it is clear that, when a resolution over a complete set of
momentum states $\left| \mathbf{q}^{\prime }\right\rangle $ is inserted
between $\mathcal{T}_{o}^{\dagger }$ and $\mathcal{T}_{o}$ and the
expression (\ref{Tnought}) for the matrix elements of $\mathcal{T}_{o}$ is
used, the diagonal elements $\left\langle \mathbf{q}|\mathcal{A}|\mathbf{q}%
\right\rangle $ involve the \textit{square }of Dirac delta functions $\delta
(q-q^{\prime })$. To evaluate these the ``magnitude'' of $\delta (0)$ must
be somehow set. This is done by relating it to an original normalization
volume of the box. While not implausible, such a protocol is certainly not
rigorous and is open to question.

To avoid the necessity of this kind of maneuver we will employ bath states
$\left| \psi \right\rangle $ that are normalized and \emph{localized}, as is
required by a strict application of scattering theory. Before addressing the
full calculation for a bath in thermal equilibrium we consider scattering
involving a single state $\left| \psi \right\rangle $.

\subsection{Scattering of a single bath ket}

From the equations (\ref{momentumdecomp}) and (\ref{A21})
for $\left\langle \psi |\mathcal{A}%
|\psi \right\rangle $ in terms of $\left\langle \mathbf{q}_{2}|\mathcal{A}|%
\mathbf{q}_{1}\right\rangle $  it is clear that we
require integrals of the form
\begin{eqnarray}
I_{1} &=&\int d\mathbf{q}_{1}d\mathbf{q}_{2}u(\mathbf{q}_{1},\mathbf{q}%
_{2})\left\langle \mathbf{q}_{2}|\mathcal{T}_{o}+\mathcal{T}_{o}^{\dagger }|%
\mathbf{q}_{1}\right\rangle ,  \label{Idef} \\
I_{2}(\mathbf{R)} &=&\int d\mathbf{q}_{1}d\mathbf{q}_{2}u(\mathbf{q}_{1},%
\mathbf{q}_{2})\left\langle \mathbf{q}_{2}|\mathcal{T}_{o}^{\dagger }e^{i%
\mathfrak{p}\cdot \mathbf{R}/\hbar }\mathcal{T}_{o}|\mathbf{q}_{1}\right\rangle ,
\nonumber
\end{eqnarray}
which we work out in Appendix \ref{sec:app1} for an arbitrary function $u(\mathbf{q}_{1},%
\mathbf{q}_{2})$ of the two momentum variables. We find that we can
write these expressions exactly as
\begin{equation}
I_{1}=\int d\mathbf{q}\int_{\mathbf{\hat{q}}^{\perp }}d\mathbf{\Delta \,}u(%
\mathbf{q-}\frac{\mathbf{\Delta }}{2},\mathbf{q+}\frac{\mathbf{\Delta }}{2}%
)M_{1}(\mathbf{q,\Delta })  \label{I1result}
\end{equation}
and
\begin{equation}
I_{2}(\mathbf{R)=}\int d\mathbf{\hat{n}\,}d\mathbf{q}\int_{\mathbf{\hat{q}}%
^{\perp }}d\mathbf{\Delta \,}u(\mathbf{q-}\frac{\mathbf{\Delta }}{2},\mathbf{%
q+}\frac{\mathbf{\Delta }}{2})e^{i\mathbf{Q\cdot R}/\hbar }M_{2}(\mathbf{q,%
\hat{n},\Delta ),}  \label{I2result}
\end{equation}
The integration over $\mathbf{q}$ covers  all momentum space, while $\mathbf{%
\Delta }$ is a two dimensional momentum vector ranging over the plane
perpendicular to $\mathbf{q}$;  $\mathbf{\hat{n}}$ is a unit vector with $%
d\mathbf{\hat{n}}$ the associated solid angle element. Moreover,
\begin{equation}
M_{1}(\mathbf{q,\Delta })=\frac{1}{2\pi \hbar q}\left( f(\mathbf{q+}\frac{%
\mathbf{\Delta }}{2},\mathbf{q-}\frac{\mathbf{\Delta }}{2})+f^{*}(\mathbf{q-}%
\frac{\mathbf{\Delta }}{2},\mathbf{q+}\frac{\mathbf{\Delta }}{2})\right)
\label{M1def}
\end{equation}
and
\begin{equation}
M_{2}(\mathbf{q,\hat{n},\Delta )=}\frac{1}{4\pi ^{2}\hbar ^{2}}\frac{Q}{q}%
f^{*}(\mathbf{Q},\mathbf{q+}\frac{\mathbf{\Delta }}{2})f(\mathbf{Q},\mathbf{%
q-}\frac{\mathbf{\Delta }}{2}).  \label{M2def}
\end{equation}
with
\begin{equation}
\mathbf{Q}=\mathbf{\hat{n}}\sqrt{q^{2}+\frac{\Delta ^{2}}{4}}\;.  \label{Qdef}
\end{equation}
With these formulas in hand we can address the expression for $\left\langle
\psi |\mathcal{A}|\psi \right\rangle $ once $\left| \psi \right\rangle $ is
specified. To do this, we take the bath particle wave function $\left\langle
\mathbf{r}^{\prime }|\psi \right\rangle $ to be a Gaussian wave packet
centered at $\mathbf{r}_{o}$ in position and $\mathbf{p}_{o}$ in momentum,
\[
\left\langle \mathbf{r}^{\prime }|\psi \right\rangle =\frac{e^{i\mathbf{p}%
_{o}\mathbf{\cdot }\left( \mathbf{r}^{\prime }-\mathbf{r}_{o}\right) /\hbar }%
}{\left( \pi a^{2}\right) ^{3/4}}e^{-\left| \mathbf{r}^{\prime }-\mathbf{r}%
_{o}\right| ^{2}/(2a^{2})},
\]
and characterized by
\begin{eqnarray*}
\Delta x &=&\frac{a}{\sqrt{2}}, \\
\Delta p_{x} &=&\frac{b}{\sqrt{2}},
\end{eqnarray*}
with $ab=\hbar $. For this minimum uncertainty wave packet we find that the
expression for $\left\langle \psi |\mathbf{q}_{2}\right\rangle \left\langle
\mathbf{q}_{1}|\psi \right\rangle $ in the integral (\ref{momentumdecomp})
for $\left\langle \psi |\mathcal{A}|\psi \right\rangle $ becomes
\[
\big\langle \psi \big|\mathbf{q}+\frac{\mathbf{\Delta }}{2}\big\rangle
\big\langle \mathbf{q}-\frac{\mathbf{\Delta }}{2}\big|\psi\big \rangle
=\left( \frac{1}{\pi b^{2}}\right) ^{3/2}e^{i\mathbf{\Delta }\cdot \mathbf{r}%
_{o}/\hbar }e^{-\Delta ^{2}/(4b^{2})}e^{-\left| \mathbf{q}-\mathbf{p}%
_{o}\right| ^{2}/b^{2}}.
\]
\begin{figure}
 \centering
  \includegraphics[width=0.66\linewidth]{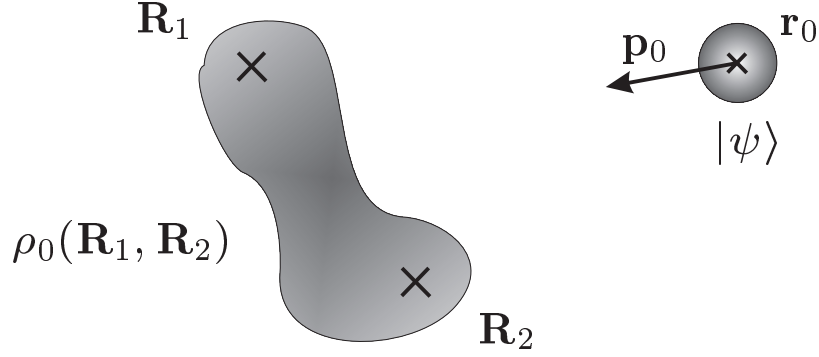}
  \caption{For this configuration the direct product of $\rho_{o}$ and $|\psi\rangle\langle\psi|$
  can be taken as both the total asymptotic-in density operator  and
  the initial density operator at $t=0$.}
  \label{fig:2}
\end{figure}
We now assume that this wave packet is located far enough away from the
regions of space where an initial density operator (\ref{rhonought}) is
concentrated, and with an average momentum directed towards the Brownian
particle such that the combined density operator (\ref{rhoin}) can be taken
both as an initial density operator at $t=0$, and as the asymptotic-in
density operator (see Fig.~2). Then using the expressions above we find
\begin{equation}
\left\langle \psi |\mathcal{A}|\psi \right\rangle =\left( \frac{1}{\pi b^{2}}%
\right) ^{3/2}\int d\mathbf{\hat{n}\,}d\mathbf{q\;}e^{-\left| \mathbf{q}-%
\mathbf{p}_{o}\right| ^{2}/b^{2}}\int_{\mathbf{\hat{q}}^{\perp }}d\mathbf{%
\Delta \,}\mathcal{B}(\mathbf{\hat{n},q,\Delta )}
\;,
\label{inttodo}
\end{equation}
where
\begin{eqnarray*}
\mathcal{B}(\mathbf{\hat{n},q,\Delta )} &=&e^{i\mathbf{q\cdot }\left(
\mathbf{R}_{1}-\mathbf{R}_{2}\right) /\hbar }e^{i\mathbf{Q\cdot (R}_{2}-%
\mathbf{R}_{1})/\hbar }e^{-i\mathbf{\Delta \cdot (r}_{o}\mathbf{-\overline{R}}%
)/\hbar }e^{-\Delta ^{2}/(4b^{2})}M_{2}(\mathbf{q,\hat{n},\Delta )} \\
&&-\frac{1}{2}e^{-i\mathbf{\Delta \cdot (r}_{o}\mathbf{-R}_{1})/\hbar
}e^{-\Delta ^{2}/(4b^{2})}M_{2}(\mathbf{q,\hat{n},\Delta )} \\
&&-\frac{1}{2}e^{-i\mathbf{\Delta \cdot (r}_{o}\mathbf{-R}_{2})/\hbar
}e^{-\Delta ^{2}/(4b^{2})}M_{2}(\mathbf{q,\hat{n},\Delta )} \\
&&+\frac{i}{2}\left[ e^{-i\mathbf{\Delta }\cdot (\mathbf{r}_{o}\mathbf{-R}%
_{1})/\hbar }-e^{-i\mathbf{\Delta }\cdot (\mathbf{r}_{o}\mathbf{-R}%
_{2})/\hbar }\right] e^{-\Delta ^{2}/(4b^{2})}M_{1}(\mathbf{q,\Delta }),
\end{eqnarray*}
and where we have put
\begin{equation}
\mathbf{\overline{R}}\equiv \frac{\mathbf{R}_{1}+\mathbf{R}_{2}}{2}.
\label{Rmdef}
\end{equation}
The Gaussian functions will keep $\Delta $ within about $b$ of zero and $%
\mathbf{q}$ within about $b$ of $\mathbf{p}_{o}\mathbf{.}$ We now assume
that the central momentum $\mathbf{p}_{o}$ is much greater in magnitude than
its variance, $p_{o}\gg b$, and hence $q\gg b$ for all $q$ that make a
significant contribution; we also assume that the scattering amplitude varies
little over the momentum range $b$. Then we can put
\begin{eqnarray*}
M_{1}(\mathbf{q,\Delta }) &\approx &\frac{1}{\pi \hbar q}\mathrm{Re}\left( f(%
\mathbf{q},\mathbf{q})\right)  \\
M_{2}(\mathbf{q,\hat{n},\Delta )} &\approx &\frac{1}{4\pi ^{2}\hbar ^{2}}%
\left| f(q\mathbf{\hat{n}},\mathbf{q})\right| ^{2}.
\end{eqnarray*}
Once these approximation are made the integral over $\mathbf{\Delta }$ of
the three terms in $\mathcal{B}(\mathbf{\hat{n},q,\Delta )}$ can be done
immediately. The integral over $\mathbf{\Delta }$ of the first term is not
so simple because $\Delta $ still appears in $\mathbf{Q}$. In the
exponential we have phase factors that vary as
\begin{eqnarray*}
\frac{\mathbf{Q\cdot (R}_{2}-\mathbf{R}_{1})}{\hbar } &=&\sqrt{q^{2}+\frac{%
\Delta ^{2}}{4}}\frac{\mathbf{\hat{n}\cdot (R}_{2}-\mathbf{R}_{1})}{\hbar }
\\
&=&\frac{q\mathbf{\hat{n}\cdot (R}_{2}-\mathbf{R}_{1})}{\hbar }+\frac{\Delta
^{2}\mathbf{\hat{n}\cdot (R}_{2}-\mathbf{R}_{1})}{8\hbar q}+...
\end{eqnarray*}
The first correction term is of order
\begin{equation}
\frac{b^{2}\left| \mathbf{R}_{2}-\mathbf{R}_{1}\right| }{\hbar q}=\frac{%
\left( \left| \mathbf{R}_{2}-\mathbf{R}_{1}\right| /a\right) }{(q/b)}
\label{neglect}
\end{equation}
Since $q\gg b$ this term will still be much smaller than unity even if the
distance between the two positions of the Brownian particle is several
widths of the wave packet. We assume that $\left| \mathbf{R}_{2}-\mathbf{R}%
_{1}\right| $ is indeed such this quantity is much less than unity. Then we can replace the phase by its
leading order expansion
\[
\frac{\mathbf{Q\cdot (R}_{2}-\mathbf{R}_{1})}{\hbar }\approx \frac{q\mathbf{%
\hat{n}\cdot (R}_{2}-\mathbf{R}_{1})}{\hbar },
\]
in the exponentials of the first two integrals, and  the
integration  over $\mathbf{\Delta }$ can be done as well. These
are two dimensional integrals over a plane perpendicular to
$\mathbf{q}$, and so they are of the form
\[
\int_{\mathbf{\hat{q}}^{\perp }}d\mathbf{\Delta }e^{-i\mathbf{\Delta \cdot (r%
}_{o}\mathbf{-\overline{R}})/\hbar }e^{-\Delta ^{2}/(4b^{2})}=(2\pi \hbar
)^{2}\Gamma _{\mathbf{q}}(\mathbf{r}_{o}\mathbf{-\overline{R})},
\]
where we have used the fact that $ab=\hbar $ and introduced
\begin{equation}
\Gamma _{\mathbf{q}}(\mathbf{R)=}\frac{\exp \left( -\left[ R^{2}-\left(
\mathbf{\hat{q}\cdot R}\right) ^{2}\right] /a^{2}\right) }{\pi a^{2}}
\label{eq:Gammadef}
\end{equation}
which involves $R^{2}-\left( \mathbf{\hat{q}\cdot R}\right) ^{2}$, the square of
the component of $\mathbf{R}$ that is perpendicular to $\mathbf{q}\equiv q\,\mathbf{\hat{q}}$.  In all
we find
\begin{equation}
\left\langle \psi |\mathcal{A}|\psi \right\rangle =\int d\mathbf{q}\frac{%
e^{-\left| \mathbf{q}-\mathbf{p}_{o}\right| ^{2}/b^{2}}}{\left( \pi
b^{2}\right) ^{3/2}}A^{\mathbf{r}_{o}}(\mathbf{q),}  \label{efguse}
\end{equation}
where
\begin{eqnarray}
A^{\mathbf{r}_{o}}(\mathbf{q)} &=&\Gamma _{\mathbf{q}}(\mathbf{r}_{o}\mathbf{%
-\overline{R})}\int d\mathbf{\hat{n}\;}e^{i(\mathbf{q-}q\mathbf{\hat{n}})\mathbf{%
\cdot }\left( \mathbf{R}_{1}-\mathbf{R}_{2}\right) /\hbar }\left| f(q\mathbf{%
\hat{n}},\mathbf{q})\right| ^{2}  \label{fguse} \\
&&-\frac{1}{2}\left( \Gamma _{\mathbf{q}}(\mathbf{r}_{o}\mathbf{-R}_{1})%
\mathbf{+}\Gamma _{\mathbf{q}}(\mathbf{r}_{o}\mathbf{-R}_{2}\mathbf{)}%
\right) \int d\mathbf{\hat{n}\;}\left| f(q\mathbf{\hat{n}},\mathbf{q}%
)\right| ^{2}  \nonumber \\
&&+\frac{2\pi i\hbar }{q}\left( \Gamma _{\mathbf{q}}(\mathbf{r}_{o}\mathbf{-R%
}_{1})\mathbf{-}\Gamma _{\mathbf{q}}(\mathbf{r}_{o}\mathbf{-R}_{2}\mathbf{)}%
\right) \mathrm{Re}\left( f(\mathbf{q},\mathbf{q})\right) .  \nonumber
\end{eqnarray}

This is the result we will find most useful when we move to a thermal
distribution of bath particles. But we close this section with an observation that
is of interest in its own right. Consider the special case where the size of
the bath particle wave packet is much larger than the distance between the
points $\mathbf{R}_{1}$ and $\mathbf{R}_{2}$,
\begin{equation}
\left| \mathbf{R}_{1}-\mathbf{R}_{2}\right| /a=\left| \mathbf{R}_{1}-\mathbf{%
R}_{2}\right| b/\hbar \ll 1.  \label{strongcondition}
\end{equation}
Then in the $\Gamma _{\mathbf{q}}$ functions of (\ref{fguse}) we can replace
$\mathbf{R}_{1}$ and $\mathbf{R}_{2}$ by $\mathbf{\overline{R}}$. Moreover, since
the integral in (\ref{efguse}) restricts $\mathbf{q}$ to within a distance
of about $b$ of $\mathbf{p}_o$, in (\ref{fguse}) we can replace $\mathbf{q}$
by $\mathbf{p}_o$ in the scattering amplitudes and in the phase, using the
assumption already made that they vary little over a range of $b$; we can
also replace the $\Gamma _{\mathbf{q}}$ functions by corresponding $\Gamma _{%
\mathbf{p}_o}$ functions. The integral in (\ref{efguse}) can then be done, and
using (\ref{deltarho}) we find
\[
\Delta \rho (\mathbf{R}_{1}\mathbf{,R}_{2})=-\rho _{o}(\mathbf{R}_{1}\mathbf{%
,R}_{2})\Gamma _{\mathbf{p}_o}(\mathbf{r}_{o}\mathbf{-\overline{R})}\int d\mathbf{%
\hat{n}}\left( 1-e^{i(\mathbf{p}_o-p_o\mathbf{\hat{n})\cdot (R}_{1}-\mathbf{R}%
_{2})/\hbar }\right) \left| f(p_o\mathbf{\hat{n}},\mathbf{p}_o)\right| ^{2}.
\]
The physics here is transparent since
\[
\Gamma _{\mathbf{p}_o}(\mathbf{r}_{o}\mathbf{-\overline{R})=}\frac{e^{-\ell^{2}/a^{2}}%
}{\pi a^{2}}
\;,
\]
where $\ell$ is the impact parameter of the collision (see
Fig.~3). Decoherence occurs only if the bath particle does not
``miss'' the Brownian particle, \textit{i.e., }if the impact
parameter $\ell$ is smaller than the  wave packet extension (as
limited by the uncertainty principle); its maximum effect scales
with the integrated square of the amplitude of the normalized
bath particle as it passes over the pair of points ($1/(\pi
a^{2})$); it vanishes as $\left|
\mathbf{R}_{1}-\mathbf{R}_{2}\right| \rightarrow 0$ because the
scattering by the two points then becomes identical; and it
depends only on the scattering amplitude at momentum magnitude
$p_o$
because the variation of that scattering amplitude over
the range of momentum components included in the wave packet has
been neglected.

\begin{figure}
 \centering
  \includegraphics[width=0.66\linewidth]{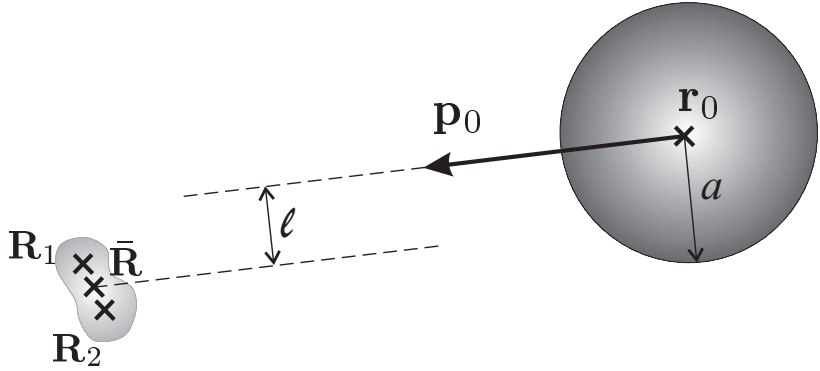}
  \caption{A configuration where $a\gg R=|\mathbf{R}_1-\mathbf{R}_2|$;
  $\ell=\sqrt{u^2-(\mathbf{u}\cdot\mathbf{\hat{p}_o})^2}$ is
  the impact parameter, where $\mathbf{u}=\mathbf{r}_{o}-\bar{\mathbf{R}}$.}
  \label{fig:3}
\end{figure}

\subsection{Convex decompositions of the bath density operator}

To apply the results derived above to a Brownian particle subject
to a thermal bath we need to describe the effect of the thermal
bath in terms of incident, normalized wave packets. We start with
a single bath particle restricted to a normalization volume
$\Omega$.
In thermal equilibrium, at temperature $k_{B}T=\beta ^{-1}$,
the bath state is specified by the
density operator
\begin{equation}
\rho ^{\rm bath}=\frac{\lambda ^{3}}{\Omega }
e^{-\beta {\mathfrak{p}}^2/(2m)}
%e^{-\beta H_{\rm bath}}
\label{rhoexpression}
\end{equation}
provided $\Omega$ is much larger than the cube of the thermal de
Broglie wave length
\begin{equation}
\lambda =\sqrt{\frac{2\pi \hbar ^{2}\beta }{m}}\;.
\label{deBroglie}
\end{equation}

The usual convex decomposition of (\ref{rhoexpression}) in terms
of the delocalized energy eigenstates is the obvious one and,
aside from the freedom in choosing orthogonal states from among a
degenerate set, it is the only one in terms of orthogonal states.
But a host of others can also found. A particularly convenient
set of convex decompositions for our problem at hand can be
obtained by using
\begin{equation}
e^{-\beta {\mathfrak{p}}^2/(2m)}
%e^{-\beta H_{\rm bath}}
=\left( \frac{\bar{\beta}}{\beta }\right) ^{3/2}\int d\mathbf{p}\,
\frac{e^{-\hat{\beta}\mathbf{p}^{2}/(2m)}}{\left( 2\pi m/\hat{%
\beta}\right) ^{3/2}}\,e^{-\bar{\beta}\left( \mathfrak{p}-\mathbf{p}\right)
^{2}/(2m)}  \label{expansion}
\end{equation}
which holds as long as $\hat{\beta}$ and $\bar{\beta}$ are both positive and
\[
\frac{1}{\beta }=\frac{1}{\bar{\beta}}+\frac{1}{\hat{\beta}}
\]
or, in terms of the pseudo-temperatures $k_{\rm B}\hat{T}\equiv
\hat{\beta}^{-1}$ and $k_{\rm B}\bar{T}=\bar{\beta}^{-1}$,
\[
T=\bar{T}+\hat{T}.
\]
In order to use the decomposition (\ref{expansion})  for $\rho^{\rm bath} $,
we write
\[
e^{-\bar{\beta}\left( \mathfrak{p}-\mathbf{p}\right) ^{2}/(2m)}=e^{-\bar{\beta}%
\left( \mathfrak{p}-\mathbf{p}\right) ^{2}/(4m)}\,\mathcal{I}_\Omega\,e^{-\bar{\beta}%
\left( \mathfrak{p}-\mathbf{p}\right) ^{2}/(4m)},
\]
and take the identity operator in position representation
\[
\mathcal{I}_\Omega=\int_{\Omega} d\mathbf{r}\left| \mathbf{r}\right\rangle \left\langle
\mathbf{r}\right| ,
\]
where the integration covers the bath volume $\Omega$. We find
\begin{equation}
\rho ^{\rm bath}=\int_\Omega \frac{d\mathbf{r}}{\Omega
}\int d\mathbf{p}\,\hat{\mu}(\mathbf{p})\mathbf{\,}
\,\left| \psi _{\mathbf{rp}}\right\rangle \left\langle \psi _{%
\mathbf{rp}}\right| ,  \label{rhobuse}
\end{equation}
where
\begin{equation}
\hat{\mu}(\mathbf{p})=\left( \frac{\hat{\beta}}{2\pi m}\right) ^{3/2}
e^{-\hat{\beta}\mathbf{p}^{2}/(2m)}
\label{muhat}
\end{equation}
is {a} normalized momentum distribution function,
$\int \hat{\mu}(\mathbf{p})d\mathbf{p}=1$, and the states
\begin{eqnarray}
\left| \psi _{\mathbf{rp}}\right\rangle &\equiv &\lambda ^{3/2}\left( \frac{%
\bar{\beta}}{\beta }\right) ^{3/4}e^{-\bar{\beta}\left( \mathfrak{p}-\mathbf{p}%
\right) ^{2}/(4m)}\left| \mathbf{r}\right\rangle  \label{psidef} \\
&=&\bar{\lambda}^{3/2}e^{-\bar{\beta}\left( \mathfrak{p}-\mathbf{p}\right)
^{2}/(4m)}\left| \mathbf{r}\right\rangle ,  \nonumber
\end{eqnarray}
are characterized by the length scale
\[
\bar{\lambda}=\sqrt{\frac{2\pi \hbar ^{2}\bar{\beta}}{m}}
\;,
\]
(compare with (\ref{deBroglie})). One then immediately finds
\begin{equation}
\left\langle \mathbf{r}^{\prime }|\psi _{\mathbf{rp}}\right\rangle =\frac{2%
\sqrt{2}}{\bar{\lambda}^{3/2}}e^{i\mathbf{p\cdot }\left( \mathbf{r}^{\prime
}-\mathbf{r}\right) /\hbar }e^{-2\pi \left| \mathbf{r}^{\prime }-\mathbf{r}%
\right| ^{2}/\bar{\lambda}^{2}}  \label{psiform}
\end{equation}
so the wave packet $|\psi _{\mathbf{rp}}\rangle $ is centered at
$\mathbf{r}$ and has an average momentum $\mathbf{p}$. Indeed,
it is of the Gaussian form used in the preceding section with
minimal uncertainties,
$b\equiv \sqrt{2mk_{\rm B}\bar{T}}$ and $a\equiv \hbar /b$. Thus
these wave packets have
\[
\frac{\left( \Delta p_{x}\right) ^{2}}{2m}=\frac{k_{\rm B}\bar{T}}{2}
\;,
\]
while if we calculate the momentum variance associated with the
distribution function $\hat{\mu}(\mathbf{p})$,
\[
\left( \delta p_{x}\right) ^{2}\equiv \int p_{x}^{2}\hat{\mu}(\mathbf{p})d%
\mathbf{p}
\;,
\]
we find
\[
\frac{\left( \delta p_{x}\right) ^{2}}{2m}=\frac{k_{\rm B}\hat{T}}{2} \;.
\]
That is,
\[
\frac{\left( \Delta p_{x}\right) ^{2}}{2m}+\frac{\left( \delta p_{x}\right)
^{2}}{2m}=\frac{k_{\rm B}T}{2}
\;.
\]
We see that in the class (\ref{rhobuse},\ref{muhat}) of convex
decompositions of $\rho^{\rm bath}$ a part of the thermal kinetic
energy is associated with the size of the wave packets
themselves, while the rest resides in the motion of the centres
of the wave packets. If we take $\hat{T}\rightarrow 0$ then the
wave packets are essentially all at rest characterized by a size
$\bar{\lambda}\rightarrow \lambda $, which is the thermal de
Broglie wavelength. On the other hand, for $\hat{T}\gg \bar{T}$
the wave packets are much larger than the thermal de Broglie
wavelength, and essentially all the thermal kinetic energy is
associated with the expectation value of the momenta of the wave
packets; we have
\[
\frac{\left\langle \mathbf{p}^{2}\right\rangle }{2m}\equiv \frac{1}{2m}\int
\mathbf{p}^{2}\hat{\mu}(\mathbf{p})d\mathbf{p=}\frac{3}{2}k_{\rm B}\hat{T},
\]
and so a typical speed for the wave packets is
\[
v_{\rm wp}\equiv \sqrt{\frac{3k_{\rm B}\hat{T}}{m}}.
\]

\subsection{Scattering of a thermal bath of particles}

With the preliminaries of the preceding sections we are now in a position
to begin the calculation of the effect of a thermal bath on the coherence of
a  massive Brownian particle. We begin with a number of assumptions
and choices that  will be made, and then discuss their applicability and
relevance in the context of the calculation.

\subsubsection{Assumptions and choices}

\begin{enumerate}
\item  We neglect initial correlations, taking  the initial
full density operator  at $t=0$ to be a direct product of a
Brownian particle density operator and a density operator for the
bath particles in thermal equilibrium,
\begin{equation}
\rho ^{\rm total}(t=0)=\rho _{o}\otimes \rho ^{\rm bath}
\;.
\label{uncorrelated}
\end{equation}

\item  We assume that the density of bath particles is much less than $%
\lambda ^{-3}$; then the issue of particle degeneracy does not matter and we may
consider the density operator of the total bath to be just the product of
density operators for individual particles. Thus we can calculate effects
`particle by particle'. We choose a volume $\Omega $ much larger than any
other volume of interest.

\item  We use a convex decomposition of $\rho ^{\rm bath}$ for a single bath
particle of the type described above, with $\bar{T}\ll T$ such that
\[
\hat{T}\approx T
\]
and therefore
\begin{equation}
b^{2}\ll \left\langle \mathbf{p}^{2}\right\rangle
\;.  \label{bcondition}
\end{equation}
This renders  $b$  sufficiently small so that the variation in scattering
amplitudes over the momentum spread of a wave packet is negligible for
essentially all of the wave packets in the convex decomposition.

\item  The value of $\bar{T}$ should also be small enough that the neglect of the
variation of the scattering amplitudes in the integral (\ref{inttodo}) is
justified, and that we can use the approximation of neglecting terms on the
order of (\ref{neglect}) above. For the latter we need
\[
\frac{b^{2}}{\hbar q}\,| \mathbf{R}_{2}-\mathbf{R}_{1}|=
%\frac{\hbar \left| \mathbf{R}_{2}-\mathbf{R}_{1}\right| }{qa^{2}}\ll 1
\frac{2mk_{\rm B}\bar{T}}{\hbar q}\,| \mathbf{R}_{2}-\mathbf{R}_{1}|
\ll 1
\]
Now for typical wave packets the average momentum $p$, and hence $q$, will be of the order of
$mv_{\rm wp}=\sqrt{3mk_{\rm B}\hat{T}}$, so this condition becomes
%\[
%\frac{\hbar \left| \mathbf{R}_{2}-\mathbf{R}_{1}\right| }{mv_{\rm wp}a^{2}}=%
%\sqrt{\frac{8\pi }{3}\frac{\hat{T}}{T}}\frac{\left| \mathbf{R}_{2}-\mathbf{R}%
%_{1}\right| }{\lambda }\frac{\bar{T}}{T},
%\]
\[
\frac{2k_{\rm B}\bar{T}}{\hbar v_{\rm wp}}\,| \mathbf{R}_{2}-\mathbf{R}_{1}|
=\Big(\frac{8\pi}{3}\Big)^{1/2}\,\frac{\bar{T}}{\sqrt{T\hat{T}}}\,
\frac{| \mathbf{R}_{2}-\mathbf{R}_{1}|}{\lambda}
\ll 1
\;,
\]
and since  $\hat{T}\approx T$ this reduces to
\[
\bar{T}\ll
\frac{\left| \mathbf{R}_{2}-\mathbf{R}_{1}\right| }{\lambda }\;T
\;.
\]

\item  We choose a coarse-graining time $\Delta t$ sufficiently large that
\begin{eqnarray}
v_{\rm wp}\Delta t &\gg &a,  \label{dtconditions} \\
v_{\rm wp}\Delta t &\gg &\left| \mathbf{R}_{2}-\mathbf{R}_{1}\right| .
\nonumber
\end{eqnarray}
That is, a typical packet travels a distance much greater than its width
and much greater than the distance between the two decohering sites during
the coarse graining time. Using the expressions for $v_{\rm wp}$ and $a$ above,
and $\hat{T}\approx T$, the first condition reads
\begin{equation}
\Delta t\gg \frac{\hbar }{k_{\rm B}T}\sqrt{\frac{T}{\bar{T}}}.
\label{firstdtcondition}
\end{equation}
The second condition is essentially independent of $\bar{T}$ as long as $%
\hat{T}\approx T$, and simply demands  that a typical bath particle
wave packet can travel many times the distance between the decohering sites
during the coarse-graining time.
\end{enumerate}

It is easy to see that for any $\mathbf{R}_{1}$ and $\mathbf{R}_{2}$ of
interest we can meet all these conditions by the choice of a large enough $%
\Delta t$, and we will see below that for a small enough density the change
in the reduced density operator will be small over any given $\Delta t$.
Hence our approximations will generally be valid in the low density limit.

\subsubsection{Calculation}

\begin{figure}
 \centering
  \includegraphics[width=0.5\linewidth]{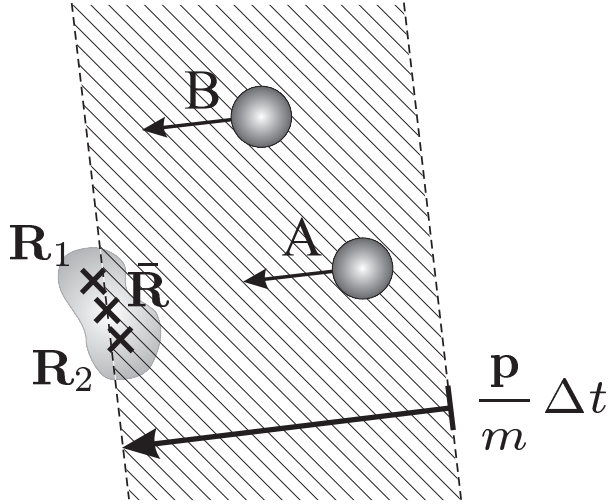}
  \caption{For wave packets with momentum $\mathbf{p}$ and centers in the
  hatched region $\mathsf{R}(\mathbf{p})$ we take the initial density
  operator to be the direct product $\rho_{o}\otimes \left|
  \psi _{\mathbf{rp}}\right\rangle \left\langle \psi _{\mathbf{rp}}\right|$.
  For some wave packets, such as (A), an actual collision will occur, while
  for others such as (B) one will not.}
  \label{fig:4}
\end{figure}

The convex decomposition (\ref{rhobuse}) we use for the density
operator of a bath particle leads us to think almost classically
about the collision of the wave packets $|\psi
_{\mathbf{rp}}\rangle $ with our Brownian particle. In a naive
classical picture the bath particles of a given
$\mathbf{p}$ that can be considered as
coming in towards a collision with the Brownian particle in time
$\Delta t$ are those that lie on any of the planes perpendicular
to $\mathbf{p}$ and extending out a distance $p\Delta t/m$ in the
direction $-\mathbf{\hat{p}}$ from $\mathbf{\overline{R}}$
(recall (\ref{Rmdef}); see Fig. 4). Of course, some of these will
completely miss the Brownian particle, but none have had a
collision with it in the past. For a given $\mathbf{p}$ we refer
to this region of space as $\mathsf{R}(\mathbf{p})$.

Returning to the wave packets, note that those with central positions $%
\mathbf{r}$ close to the $\mathbf{R}_{1}$ or $\mathbf{R}_{2}$ of interest
will initially be overlapping with regions of space for which $\rho _{o}(%
\mathbf{R}_{1},\mathbf{R}_{2})$ is non-vanishing; here any talk of a
collision is inappropriate, since at initiation, at $t=0$, the Brownian and
bath particle would immediately be strongly interacting. This is an artifact
of the assumption (\ref{uncorrelated}) of an initially uncorrelated state,
which is clearly unphysical. If the uncorrelated state were taken seriously,
there would be a rapid ``jolt,'' or shift in the reduced density operator
due to the set up of correlations \cite{WeissChap2}. These effects we neglect here, as
they are implicitly neglected in most such calculations; we do note that,
for $\Delta t$ large enough, the regions $\mathsf{R}(\mathbf{p})$ for $%
\mathbf{p}$ of interest will be sufficiently large that only a small
fraction of the wave packets in the sphere will fall in this problematic
class.

In calculations involving the rest of the wave packets in $\mathsf{R}(\mathbf{%
p})$, we can take the initial density operator also to be the asymptotic-in
density operator, and use the scattering theory calculation for a bath
particle wave packet given above. Considering $N$ bath particles in volume $%
\Omega $, the total result for $\Delta \rho (\mathbf{R}_{1}\mathbf{,R}_{2})$
(recall (\ref{deltarho}) and (\ref{rhobuse})) is then
\[
\Delta \rho (\mathbf{R}_{1}\mathbf{,R}_{2})=N\rho
_{o}(\mathbf{R}_{1},\mathbf{R}_{2})\int d
\mathbf{p\,}\hat{\mu}(\mathbf{p})\int_{\mathsf{R}(\mathbf{p})}\frac{d\mathbf{r}}{\Omega
}
\left\langle \psi _{\mathbf{rp}}|\mathcal{A}%
|\psi _{\mathbf{rp}}\right\rangle ,
\]
where we assume that the inclusion of the problematic class of
wave packets identified above will not lead to serious error.
Using the result (\ref{efguse}) from our scattering calculation
above, we have
\[
\Delta \rho (\mathbf{R}_{1}\mathbf{,R}_{2})=n\rho
_{o}(\mathbf{R}_{1},\mathbf{R}_{2})\int
d\mathbf{p\,}\hat{\mu}(\mathbf{p})\int_{\mathsf{R}(\mathbf{p})}d\mathbf{r}\,\int
d\mathbf{q}\frac{e^{-\left| \mathbf{q}-\mathbf{p}\right|
^{2}/b^{2}}}{\left( \pi b^{2}\right)
^{3/2}}A^{\mathbf{r}}(\mathbf{q).}
\]
If in fact the strong condition (\ref{strongcondition}) can be assumed then
the integral over $\mathbf{q}$ can be evaluated immediately, as was done following (%
\ref{strongcondition}), and then also the integral over $\mathbf{p}$ can be
performed. But this is not necessary. We can simply note that, by virtue of (%
\ref{bcondition}), $\hat{\mu}(\mathbf{p})$ will vary little over the range $b
$ that $e^{-\left| \mathbf{q}-\mathbf{p}\right| ^{2}/b^{2}}$peaks and falls.
Hence we can replace $\hat{\mu}(\mathbf{p})$ by $\hat{\mu}(\mathbf{q})$ and $%
\mathsf{R}(\mathbf{p})$ by $\mathsf{R}(\mathbf{q})$, and immediately do the
integral over $\mathbf{p}$ to yield
\[
\Delta \rho (\mathbf{R}_{1}\mathbf{,R}_{2})=n\rho _{o}(\mathbf{R}_{1},%
\mathbf{R}_{2})\int d\mathbf{q\,}
\hat{\mu}(\mathbf{q})\int_{\mathsf{R}(\mathbf{q})}d\mathbf{r}\,A^{\mathbf{r}}(\mathbf{q).}
\]
Since the only $\mathbf{r}$ dependence is in the
$\Gamma_\mathbf{q}$, see (\ref{eq:Gammadef}), one can now do the
$\mathbf{r}$ integral for each fixed $\mathbf{q}$, putting
$d\mathbf{r=}d\mathbf{r}^{\bot }dr^{\Vert }$, where $r^{\Vert }$
refers to the distance in the direction $\mathbf{-q}$. Since  the
integration over $\mathbf{r}^{\bot }$ is unrestricted in the region $\mathsf{R}(\mathbf{q})$ we
have
\[
\int_{\mathsf{R}(\mathbf{q})}d\mathbf{r}^{\bot }\Gamma
_{\mathbf{q}}(\mathbf{r}^{\bot }-\mathbf{R}^{\bot}_{i})=1
\]
for $\mathbf{R}_{i}$ $=\mathbf{R}_{1}$, $\mathbf{R}_{2}$, or $\mathbf{\overline{R}%
}$, see (\ref{eq:Gammadef}). On the other hand there is no
dependence on $r^\Vert$,
\[
\int_{\mathsf{R}(\mathbf{q})}dr^{\Vert }=\frac{q}{m}\Delta t,
\]
and so we find
\[
\frac{\Delta \rho (\mathbf{R}_{1}\mathbf{,R}_{2})}{\Delta t}=n\rho _{o}(%
\mathbf{R}_{1},\mathbf{R}_{2})\int d\mathbf{q\,}\frac{q}{m}\hat{\mu} (\mathbf{q}%
)\,\int d\mathbf{\hat{n}}\left( e^{i(\mathbf{q-}q\mathbf{\hat{n})\cdot (R}%
_{1}-\mathbf{R}_{2})/\hbar }-1\right) \left| f(q\mathbf{\hat{n}},\mathbf{q}%
)\right| ^{2}
\;.
\]
Finally,  we recall that $\hat{T}\approx T$ and therefore  put $\hat{\mu}(\mathbf{q}%
)\approx \mu (\mathbf{q})$, where
\begin{equation}
\mu (\mathbf{q})=\left(\frac{\beta}{2\pi m}\right) ^{3/2} \,
e^{-\beta \mathbf{q}^{2}/(2m)}
\;,
\label{eq:mu}
\end{equation}
\textit{cf. }equation (\ref{muhat}).
Now if $\nu(q)$ is the thermal distribution function for the
momentum magnitude of the bath particles and
$\mathbf{q}=q\mathbf{\hat{s}}$, where $\mathbf{\hat{s}}$
is a unit vector and $d\mathbf{\hat{s}}$ the associated element of solid
angle, we have
\begin{equation}
\mu (\mathbf{q})d\mathbf{q}=\frac{\nu (q)dqd\mathbf{\hat{s}}}{4\pi },
\label{muandnu}
\end{equation}
and hence on a coarse grained time scale we find
(\ref{drhodt},\ref{gamma}) with $\varepsilon=1$.

%%%%%%%%%%%%%%%%%%%%%%%%%%%%%%%%%%%%%%%%%%%%%%%%%%%%%%%%%%%%%%%%%5
%%%%%%
%%%%%%  SECTION 3
%%%%%%
%%%%%%%%%%%%%%%%%%%%%%%%%%%%%%%%%%%%%%%%%%%%%%%%%%%%%%%%%%%%%%%%%5

\section{The traditional approach: a remedy}
\label{sec:dsquared}

We showed in the preceding section how the problem of evaluating
a squared Dirac function can be circumvented by expressing the
thermal state of the bath particles in an over-complete,
non-orthogonal basis of Gaussian wave packets (see Eq.
(\ref{rhobuse})). However, it is certainly reasonable to explore
the possibility of using the standard diagonal representation of
the thermal bath density operator, which facilitates the formal
calculation considerably. After all, all the representations  of
$\rho^{\rm bath}$ are equally valid and should yield the same
master equation provided the calculation is done in a correct
way. It is therefore worthwhile to search for a way to deal
properly  with such an ill-defined object as the ``square'' of a
delta distribution function.

In this section we show how a proper evaluation of the diagonal momentum
basis matrix elements can be implemented. This leads to an alternate
derivation of the master equation (\ref{drhodt},\ref{gamma}),
and allows us to highlight the
origin of the problem plaguing earlier workers and to discuss further
implications. However, rather than attempting a mathematically rigorous
formulation, we base our presentation on a simple physical argument. Our
point is that such an argument can lead to a prescription for correctly
evaluating improper products of Dirac delta functions,
although this differs from previous naive treatments.

\subsection{A single collision}

Let us consider again the action of a \emph{single} scattering
event on the
Brownian particle in position representation
$\rho_{o}(\mathbf{R}_1,\mathbf{R}_2)$ and in the limit of a large
mass.
It follows from the discussion
in section \ref{sec:scatt}
that after the collision it differs
merely by a factor
from the initial Brownian state,
\begin{equation}
\label{eq:pps}
\rho(\mathbf{R}_1,\mathbf{R}_2)
=\eta(\mathbf{R}_1,\mathbf{R}_2)\rho_{o}(\mathbf{R}_1,\mathbf{R}_2)
%\;.
\end{equation}
which is given by
\begin{equation}
\label{eq:etadef}
\eta(\mathbf{R}_1,\mathbf{R}_2)=
\mathrm{tr}_{\rm bath}\{
e^{-i\mathfrak{p}\mathbf{R}_2/\hbar}
\mathcal{S}^\dagger_{o} e^{i\mathfrak{p}(\mathbf{R}_2-\mathbf{R}_1)/\hbar}
\mathcal{S}_{o} e^{i\mathfrak{p}\mathbf{R}_1/\hbar}
\rho^{\rm bath}\}
\;,
\end{equation}
(see Eqs. (\ref{eq:rhochange}) and (\ref{eq:overlap})). In section
\ref{sec:scatt} only  pure states $\rho^{\rm
bath}=|\psi\rangle\langle\psi|$ of the bath particle were
considered, but the reasoning is immediately generalized to  mixed
states.

The factor $\eta(\mathbf{R}_1,\mathbf{R}_2)$ may be called the
decoherence function, since it describes the effective loss
of coherence in the
Brownian state which arises from disregarding the scattered bath particle.
The normalization of $\rho^{\rm bath}$
implies
\begin{equation}
\label{eq:lim1}
\lim_{|\mathbf{R}_1-\mathbf{R}_2|\to 0}\eta(\mathbf{R}_1,\mathbf{R}_2)=1
\end{equation}
which means that the collision does not change the position
distribution of the Brownian point particle,
$\rho(\mathbf{R},\mathbf{R})=\rho_{o}(\mathbf{R},\mathbf{R})$. On
the other hand, possible quantum correlations between
increasingly far separated  points will vanish, since a collision
may be viewed as a position measurement of the Brownian particle
by the bath which destroys superpositions of distant locations:
\begin{equation}
\label{eq:lim2}
\lim_{|\mathbf{R}_1-\mathbf{R}_2|\to\infty}\eta(\mathbf{R}_1,\mathbf{R}_2)=0
\end{equation}
This complete loss of coherence implies that the collision took
place with a probability of one. It could be realized, in particular,
by taking the incoming bath particle state to be a momentum
eigenket in a box centered on one of the scattering sites.

In thermal equilibrium the density operator (\ref{rhoexpression}) of the
bath particle can be written as
\begin{eqnarray}
\rho ^{bath} &=&\frac{\lambda ^{3}}{\Omega }\sum_{\mathbf{p\in }\mathbb{P}%
_{\Omega }}e^{-\beta p^{2}/(2m)}\widetilde{\left| \mathbf{p}\right\rangle }%
\widetilde{\left\langle \mathbf{p}\right| }  \nonumber\\
&=&\frac{\left( 2\pi \hbar \right) ^{3}}{\Omega }\sum_{\mathbf{p\in }
\mathbb{P}_{\Omega }}\mu (\mathbf{p})\,\widetilde{\left| \mathbf{p}\right\rangle }%
\widetilde{\left\langle \mathbf{p}\right| },
%\label{fortysix}
\label{eq:rhoB}
\end{eqnarray}
with the normalized momentum distribution function (\ref{eq:mu}) at $\beta
=1/(k_{B}T) $. The $\widetilde{\left| \mathbf{p}\right\rangle }$ are
momentum eigenkets
normalized with respect to the bath volume $\Omega $,
\begin{equation}
\widetilde{\left| \mathbf{p}\right\rangle }=\frac{\left( 2\pi \hbar \right)
^{3/2}}{\Omega ^{1/2}}\left| \mathbf{p}\right\rangle ,
\label{eq:ptildedef}
%\label{fortyseven}
\end{equation}
and the sum involves those momenta $\mathbf{p\in }\mathbb{P}_{\Omega }$
whose associated wave functions satisfy periodic boundary
conditions on the box $\Omega$. The kets (\ref{eq:ptildedef})
form an ortho-normal basis,
\begin{equation}
\sum_{\mathbf{p\in }\mathbb{P}_{\Omega }}\widetilde{\left| \mathbf{p}%
\right\rangle }\widetilde{\left\langle \mathbf{p}\right| }=\mathcal{I}%
_{\Omega }
\;,
\label{eq:discres}
\end{equation}
where $\mathcal{I}_{\Omega }$ is the identity operator in the space
of wave functions which are periodic on $\Omega $; they must be distinguished
from the standard momentum kets
\begin{equation}
\left\langle \mathbf{r|p}\right\rangle =\frac{e^{i\mathbf{p\cdot r}/\hbar }}{%
(2\pi \hbar )^{3/2}}
\;,
\label{eq:rp}
\end{equation}
which satisfy
\[
\langle \mathbf{p|p}^{\prime }\rangle =\delta (\mathbf{p-p}^{\prime })
\]
and
span
%resolve
the full space,
\begin{equation}
\int d\mathbf{p}\left| \mathbf{p}\right\rangle \left\langle \mathbf{p}%
\right| =\mathcal{I}\text{\;.}
\label{eq:pres}
\end{equation}

Since the bath state (\ref{eq:rhoB}) is diagonal in the momentum
representation, an explicit expression for the decoherence function
(\ref{eq:etadef}) is
readily obtained:
\begin{eqnarray}
\eta (\mathbf{R}_{1},\mathbf{R}_{2}) &\rightarrow &\int d\mathbf{p}\,\mu (%
\mathbf{p})\,\widetilde{\left\langle \mathbf{p}\right| }e^{-i\frak{p}\cdot
\mathbf{R}_{2}/\hbar }\mathcal{S}_{o}^{\dagger }e^{i\frak{p\cdot }\left(
\mathbf{R}_{2}-\mathbf{R}_{1}\right) /\hbar }\mathcal{S}_{o}e^{i\frak{p}%
\cdot \mathbf{R}_{1}/\hbar }\widetilde{\left| \mathbf{p}\right\rangle }
\nonumber \\
 &= & \int d\mathbf{p}\,\mu (\mathbf{p})
 \left[ 1-
 \widetilde{\langle \mathbf{p}| }\mathcal{T}_{o}^{\dagger}
 \mathcal{T}_{o}\widetilde{|\mathbf{p}\rangle}
 +e^{i\mathbf{p}\cdot(\mathbf{R}_{1}-\mathbf{R}_{2}) /\hbar }
 \widetilde{\langle \mathbf{p}| }\mathcal{T}_{o}^{\dagger}
 e^{i\frak{p}\cdot(\mathbf{R}_{2}-\mathbf{R}_{1}) /\hbar }
 \mathcal{T}_{o}\widetilde{|\mathbf{p}\rangle}
 \right]
\nonumber \\
&= &\int d\mathbf{p}\,\mu (\mathbf{p})\left[ 1-\frac{(2\pi \hbar
)^{3}}{\Omega }\int d\mathbf{p}^{\prime }\left( 1-e^{i\left( \mathbf{p-p}%
^{\prime }\right) \cdot \left( \mathbf{R}_{1}-\mathbf{R}_{2}\right) /\hbar
}\right) \left| \left\langle \mathbf{p}^{\prime }|\mathcal{T}_{o}|\mathbf{p}%
\right\rangle \right| ^{2}\right]
\label{eq:eta2}
\end{eqnarray}
In the first line
the sum over momenta $\mathbf{p\in }\mathbb{P}_{\Omega }$ was replaced by an integral
according to the usual prescription
\[
\frac{\left( 2\pi \hbar \right) ^{3}}{\Omega }
\sum_{\mathbf{p\in }\mathbb{P}_{\Omega }}
\;\rightarrow\; \int d\mathbf{p}
\;.
\]
In the second line we introduced the operator  $\mathcal{T}_{o}=i(1-\mathcal{S}_{o})$
and used the unitarity of $\mathcal{S}_{o}$,
\[
i(\mathcal{T}_{o}-\mathcal{T}_{o}^{\dagger})=-\mathcal{T}_{o}^{\dagger}\mathcal{T}_{o}
\;,
\]
as in section \ref{sec:scatt} and
in \cite{Gallis1990a,Giulini1996a}.
The last line follows after
inserting a complete set of states (\ref{eq:pres}) and noting the relation (\ref{eq:ptildedef}).

The expression in square brackets in (\ref{eq:eta2}) should be well-defined and
finite. However, it involves two arbitrarily large quantities,
the ``quantization volume'' $\Omega$, which stems from the
normalization of the bath particle, and the squared amplitude of
the $\mathcal{T}_{o}$-operator with respect to (improper)
momentum kets.
The simple matrix element is given by  the expression
(\ref{Tnought})
\begin{equation}
\label{eq:Tme}
\langle\mathbf{p}'|\mathcal{T}_{o}|\mathbf{p}\rangle=\frac{\delta(p-p')}{2\pi\hbar p}
 \, f(\mathbf{p}',\mathbf{p})
\end{equation}
involving the scattering amplitude and a delta-function which
ensures the conservation of energy during an elastic collision.
The squared modulus of (\ref{eq:Tme}) is not well defined in
the sense of distributions, but depends on the specific limiting process from
which the Dirac delta function originates. Yet one would naturally expect
that an appropriate replacement has the form
\begin{equation}
\label{eq:Tme2}
\big|\langle\mathbf{p}'|\mathcal{T}_{o}|\mathbf{p}\rangle\big|^2
\to\delta(p-p')\,g(\mathbf{p})\,
\big|f(\mathbf{p}',\mathbf{p})\big|^2
\end{equation}
with $g(\mathbf{p})$ a function involving the quantization volume.

We note that  (\ref{eq:eta2}) displays  the correct limiting
behavior (\ref{eq:lim1})  since
the expression in round brackets vanishes as $\mathbf{R}_1\to\mathbf{R}_2$. On the other hand, for
large separations
%$|\mathbf{R}_1-\mathbf{R}_2|$
of the points the phase in (\ref{eq:eta2}) oscillates rapidly
and  it will not contribute to the integral in the limit
$|\mathbf{R}_1-\mathbf{R}_2|\to\infty$.
Therefore the limit (\ref{eq:lim2})
 allows to specify the unknown function $g(\mathbf{p})$ in (\ref{eq:Tme2}).
One obtains
\begin{equation*}
g(\mathbf{p})=\frac{\Omega}{(2\pi\hbar)^3}\,\frac{1}{\sigma(\mathbf{p})p^2}
\end{equation*}
with
\begin{equation*}
\sigma(\mathbf{p})=\int d\mathbf{\hat{n}}\,
\big|f(p\,\mathbf{\hat{n}},\mathbf{p})\big|^2
\end{equation*}
the total cross section for scattering at momentum $\mathbf{p}$.
Formally, this means that one should treat the expression
involving the squared $\delta$-function and scattering amplitude as
\begin{equation}
\label{eq:deltasquare}
\big|\delta(p-p')
f(p'\mathbf{\hat{n}},\mathbf{p})\big|^2
\to
\frac{\Omega}{2\pi\hbar \sigma(\mathbf{p})}
\, \delta(p-p') \, \big|f(p\,\mathbf{\hat{n}},\mathbf{p})\big|^2
\;.
\end{equation}

\subsection{The master equation}

A master equation can now be derived in the same spirit as above.
In the low density limit for the bath particles, the decohering
effect of each  collision can be treated independently and the overall
decoherence in a short time interval $\Delta t$
is determined only by the mean number and type of single collisions.

For bath particles with momentum $\mathbf{p}$ the mean number of
collisions is given by $j(\mathbf{p})\sigma(\mathbf{p})\Delta t$,
where  $j(\mathbf{p})=n |\mathbf{p}|/m$ is the flux, and we
replace
$1/\Omega$ by $n$, the number density of bath particles.
Hence, the average change in the Brownian state during $\Delta t$
reads
\begin{align*}
%\label{eq:}
\frac{\rho(\mathbf{R}_1,\mathbf{R}_2)-\rho_{o}(\mathbf{R}_1,\mathbf{R}_2)}{\Delta t}
=-n\int & d\mathbf{p}\,
\mu(\mathbf{p}) \frac{p}{m} \int
d\mathbf{\hat{n}}
\big(1-e^{i(\mathbf{p}-p\,\mathbf{\hat{n}})(\mathbf{R}_1-\mathbf{R}_2)/\hbar}\big)
\nonumber\\
 &\times
\big|f(p\,\mathbf{\hat{n}},\mathbf{p})\big|^2 \rho_{o}(\mathbf{R}_1,\mathbf{R}_2)
\end{align*}
and noting (\ref{muandnu}) we have on a coarse-grained time scale which is much larger than the
typical interval between collisions
\begin{align}
%\label{eq:}
\frac{\partial}{\partial t}\rho(\mathbf{R}_1,\mathbf{R}_2)
=-F(\mathbf{R}_1-\mathbf{R}_2)\rho(\mathbf{R}_1,\mathbf{R}_2)\;.
\end{align}
with $F$ given by (\ref{gamma}), again with $\varepsilon=1$.

\subsection{Interpretation}

It is clear that the derivation of the decoherence function (\ref{eq:etadef}) does not
hold rigorously even for volume-normalized (\ref{eq:ptildedef}) momentum
states, since their amplitude is uniform in space and they cannot be
considered as asymptotic-in or asymptotic-out states. Nonetheless, the fact
that one obtains the ``correct'' master equation by using the diagonal
momentum representation (\ref{eq:rhoB}) indicates that it can be reasonable, at least
in a formal sense, to extend the applicability of (\ref{eq:etadef}) to volume-normalized
momentum eigenstates.

Then the appearance of the total cross section in the appropriate
replacement rule  (\ref{eq:deltasquare}) has a clear physical interpretation. The squared
matrix element of the $\mathcal{T}_{o}$ operator with respect to two
orthogonal proper states may be viewed as the probability for a transition
between the states due to a collision. The appropriate normalization of the
probability necessary in the limit of improper states is then effected by
the appearance of the total cross section $\sigma (\mathbf{p})$ in (\ref{eq:deltasquare}),
which is absent in the usual naive treatments of the squared delta function.

This point of view is confirmed by the fact that the rule (\ref{eq:deltasquare}),
which was derived from a simple physical
argument (\ref{eq:lim2}), implies a  conservation
condition. Integrating (\ref{eq:deltasquare}) we have
\begin{equation}
\label{eq:Tme3}
\frac{(2\pi\hbar)^3}{\Omega}
\int d\mathbf{p}'
\big|\langle\mathbf{p}'|\mathcal{T}_{o}|\mathbf{p}\rangle\big|^2
%\;\,\mbox{``=''}\;\,
\to
\int d\mathbf{p}'
\delta(p-p')\,
\,\frac{\big|f(\mathbf{p}',\mathbf{p})\big|^2}{p^2\,\sigma(\mathbf{p})}
=1
\end{equation}
and  hence, using (\ref{eq:pres}) and switching to volume-normalized states,
\begin{equation}
\widetilde{\langle\mathbf{p}|}\mathcal{T}_{o}\mathcal{T}_{o}^\dagger\widetilde{|\mathbf{p}\rangle}
\to 1
\;.
\label{eq:itsone}
\end{equation}
Inserting the identity (\ref{eq:discres}) yields
\begin{equation}
\sum_{\mathbf{p}'\in\mathbb{P}_\Omega}
\big|
\widetilde{\langle\mathbf{p}'|}
\mathcal{T}_{o}
\widetilde{|\mathbf{p}\rangle}
\big|^2
\to 1
\;.
\label{eq:boxsum}
\end{equation}
This is reminiscent of the situation of a multi-junction in
mesoscopic physics \cite{Imry1997a}, or of the scattering off a quantum graph \cite{Kottos2003a},
where one defines a transition matrix $\mathsf{T}_{mn}=|t_{mn}|^2$
which connects a finite number of incoming and outgoing channels.
There the $t_{mn}$ are the transmission amplitudes between the incoming and outgoing
states and the current conservation implies
\begin{equation*}
\sum_m \mathsf{T}_{mn}=1
\quad\quad\mbox{with}\quad\quad
\mathsf{T}_{mn}=|t_{mn}|^2
\;,
\end{equation*}
in analogy to (\ref{eq:boxsum}).

The fact that the conservation relation (\ref{eq:boxsum}) has no
meaningful equivalence  in the continuum limit $\Omega\to\infty$ is closely
connected to the difficulty of evaluating the squared scattering
amplitude in the momentum representation. It suggests that the
diagonal representation of $\rho^{\rm bath}$ can  be used in a
rigorous formulation of the master equation  only if the
transition of going from a discrete to a continuous set of bath
states is delayed until after the square of the scattering matrix
element  is evaluated. A calculation along this line, albeit in a
perturbative framework, is presented in the following section.

%%%%%%%%%%%%%%%%%%%%%%%%%%%%%%%%%%%%%%%%%%%%%%%%%%%%%%%%%%%%%%%%%5
%%%%%%
%%%%%%  SECTION 4
%%%%%%
%%%%%%%%%%%%%%%%%%%%%%%%%%%%%%%%%%%%%%%%%%%%%%%%%%%%%%%%%%%%%%%%%5

\section{Weak-coupling calculation}
\label{sec:weak}

We now consider an approach that is totally different from the
derivation in Section~\ref{sec:scatt}. Instead of performing a
scattering calculation, we obtain a master equation for the
reduced density operator from a weak coupling approximation that
is very similar to the analyses of quantum optics. Again the
assumption of a low density of bath particles will allow us to
calculate the effect of the bath particles one particle at a
time, so we begin with our Brownian particle and a single bath
particle restricted to a box of normalization volume $\Omega $.
While we will take the limit $\Omega \rightarrow \infty $ in the
course of the calculation, we can do it in such a way that
products of Dirac delta functions never appear.

In the absence of any interaction between the particles the Hamiltonian
reads
\[
H_{o}=\frac{\mathcal{P}^{2}}{2M}+\frac{\mathfrak{p}^{2}}{2m},
\]
where $m$ and $M$ are the bath and Brownian particle masses and $\mathfrak{p}$ and $%
\mathcal{P}$ are their momentum operators. The normalized
eigenstates of $H_{o}$ are direct products $\widetilde{| \mathbf{{P}}\rangle}\widetilde{|
\mathbf{{p}}\rangle}  $, where
$\widetilde{| \mathbf{{p}}\rangle} $ is given by
(\ref{eq:ptildedef}) with (\ref{eq:rp}),
and
similarly
\begin{equation}
\langle \mathbf{R\widetilde{|{P}}\rangle} =\frac{e^{i\mathbf{{P}%
\cdot R}/\hbar }}{\sqrt{\Omega }}\;.
\label{bigpbar}
\end{equation}
The values of $\mathbf{{p}}$ and $\mathbf{{P}}$ are restricted
to a discrete set,
$\mathbf{{p}},\mathbf{{P}}\in\mathbb{P}_\Omega$,
so
that the wave functions respect periodic boundary conditions. Our full
Hamiltonian is then
\[
H=H_{o}+V(\mathfrak{r-}\mathcal{R}),
\]
where $\mathfrak{r}$ and $\mathcal{R}$ are respectively the bath and Brownian
particle position operators, and $V$ describes the interaction.

In the interaction picture the full density operator evolves
according to
\begin{equation}
\rho _{I}^{\rm total}(t)=U(t)\rho _{I}^{\rm total}(0)U^{\dagger }(t),  \label{rhoIt}
\end{equation}
where $U(t)=1+iT(t),$ and
\[
T(t)=-\frac{1}{\hbar }\int_{0}^{t}H_{I}(t^{\prime })U(t^{\prime })dt^{\prime
},
\]
with
\begin{equation}
H_{I}(t)=e^{iH_{o}t/\hbar }V(\mathfrak{r-}\mathcal{R})e^{-iH_{o}t/\hbar }.
\label{HIt}
\end{equation}
Using the unitarity of the time evolution, $U^{\dagger }(t)U(t)=U(t)U^{\dagger }(t)=%
\mathcal{I}$, we find from (\ref{rhoIt}) and the definition of $T(t)$ that
\begin{eqnarray*}
\rho _{I}^{\rm total}(t)-\rho _{I}^{\rm total}(0) &=&\frac{1}{2}i\left[
T(t)+T^{\dagger }(t),\rho _{I}^{\rm total}(0)\right]  \\
&&+T(t)\rho _{I}^{\rm total}(0)T^{\dagger }(t)-\frac{1}{2}T^{\dagger
}(t)T(t)\rho _{I}^{\rm total}(0)-\frac{1}{2}\rho _{I}^{\rm total}(0)T^{\dagger
}(t)T(t).
\end{eqnarray*}
This equation is exact. The first term on the right side describes a
unitary modification to the dynamics of the density operator due to the
interaction with the bath particles; we neglect it here since it will not
lead to decoherence. For the other terms we make the standard
weak-coupling approximations \cite{Gardiner2000a}: we replace
$T(t)$ by
\[
T_{o}=-\frac{1}{\hbar }\int_{0}^{\Delta t}H_{I}(t^{\prime })dt^{\prime },
\]
assume an initially uncorrelated density operator, $\rho
_{I}^{\rm total}(0)=\rho _{o}\otimes \rho ^{\rm bath}$, (\textit{cf. }equation (\ref
{uncorrelated})), and at time $\Delta t$ trace over the bath to find the
change in the reduced density operator for the Brownian particle as
\begin{equation}
\Delta \rho_{I} =\mathrm{tr}_{\rm bath}\big\{ T_{o}\rho _{o}\rho ^{\rm bath}T_{o}^{\dagger
}-\frac{1}{2}T_{o}^{\dagger }T_{o}\rho _{o}\rho ^{\rm bath}-\frac{1}{2}\rho
_{o}\rho ^{\rm bath}T_{o}^{\dagger }T_{o}\big\} .  \label{deltarhotest}
\end{equation}
To construct $T_{o}$ we begin with a Fourier expansion of the interaction
potential, writing
\begin{equation}
V(\mathfrak{r-}\mathcal{R})=\frac{\left( 2\pi \hbar \right) ^{3}}{\Omega }\sum_{%
\mathbf{{q}}\in\mathbb{P}_\Omega}\bar{V}(\mathbf{{q}})e^{i\mathbf{{q}}\cdot
\left( \mathfrak{r-}\mathcal{R}\right) /\hbar },  \label{Vexpansion}
\end{equation}
choosing the prefactor for later convenience. Then forming (\ref{HIt}), by
inserting resolutions of unity (\ref{eq:discres}) in terms of the energy eigenstates between
the free evolution terms and the potential, we find
\begin{eqnarray}
H_{I}(t)
&=&\frac{\left( 2\pi \hbar \right) ^{3}}{\Omega }\sum_{\mathbf{{P}},%
\mathbf{{q}}\in\mathbb{P}_\Omega}\bar{V}(\mathbf{{q}}) \widetilde{|{\mathbf{P}-%
\mathbf{q}}\rangle} \widetilde{\langle \mathbf{{P}}|} \exp \left(
\frac{it}{\hbar }\frac{\left| {\mathbf{P}-\mathbf{q}}\right| ^{2}-%
{P}^{2}}{2M}\right)   \nonumber \\
&&\times \sum_{\mathbf{{p}}\in\mathbb{P}_\Omega}\widetilde{| {\mathbf{p}+\mathbf{q}}%
\rangle} \widetilde{\langle \mathbf{{p}}|} \exp \left( \frac{it}{%
\hbar }\frac{\left| {\mathbf{p}+\mathbf{q}}\right| ^{2}-{p}%
^{2}}{2m}\right) .  \label{HIresult}
\end{eqnarray}
Although this approach leads easily to considering the more general problem
of a finite mass Brownian particle, we defer that to a later communication.
Here we take the infinite mass limit for the Brownian particle by taking $%
M\rightarrow \infty $ in (\ref{HIresult}), and we can then write
\[
H_{I}(t)=\frac{\left( 2\pi \hbar \right) ^{3}}{\Omega }\sum_{\mathbf{{P%
},{q}}\in\mathbb{P}_\Omega}\tilde{\Lambda}_{\mathbf{{P}{q}}}\gamma _{\mathbf{%
{q}}}(t),
\]
where
\[
\tilde{\Lambda}_{\mathbf{{P}{q}}}=\widetilde{| {\mathbf{P}-%
\mathbf{q}}\rangle} \widetilde{\langle \mathbf{{P}}|}
\]
and
\[
\gamma _{\mathbf{{q}}}(t)=\bar{V}(\mathbf{{q}})\sum_{\mathbf{%
{p}}\in\mathbb{P}_\Omega}\widetilde{| {\mathbf{p}+\mathbf{q}}\rangle}
\widetilde{\langle \mathbf{{p}}|} \exp \left( \frac{it}{\hbar }\frac{%
\left| {\mathbf{p}+\mathbf{q}}\right| ^{2}-{p}^{2}}{2m}%
\right) .
\]
Using these in the expression (\ref{deltarhotest}) for $\Delta \rho $, we
find
\begin{eqnarray*}
\Delta \rho
&=&\frac{\left( 2\pi \hbar \right) ^{6}}{\Omega ^{2}}
\sum_{\mathbf{{P}}^{\prime },\mathbf{{q}}^{\prime },
\mathbf{{P}}^{\prime\prime },\mathbf{{q}}^{\prime
\prime }\in\mathbb{P}_\Omega}
\left( 2\tilde{\Lambda}_{\mathbf{%
{P}}^{\prime }\mathbf{{q}}^{\prime }}\rho _{o}\tilde{\Lambda}_{%
\mathbf{{P}}^{\prime \prime }\mathbf{{q}}^{\prime \prime
}}^{\dagger }-\tilde{\Lambda}_{\mathbf{{P}}^{\prime \prime }\mathbf{%
{q}}^{\prime \prime }}^{\dagger }\tilde{\Lambda}_{\mathbf{{P}}%
^{\prime }\mathbf{{q}}^{\prime }}\rho _{o}-\rho _{o}\tilde{\Lambda}_{%
\mathbf{{P}}^{\prime \prime }\mathbf{{q}}^{\prime \prime
}}^{\dagger }\tilde{\Lambda}_{\mathbf{{P}}^{\prime }\mathbf{{q}}%
^{\prime }}\right) \Upsilon (\mathbf{{q}}^{\prime },\mathbf{{q}}%
^{\prime \prime }),
\end{eqnarray*}
with
\[
\Upsilon (\mathbf{{q}}^{\prime },\mathbf{{q}}^{\prime \prime })=%
\frac{1}{2\hbar ^{2}}\int_{0}^{\Delta t}\int_{0}^{\Delta t}\mathrm{tr}%
_{\rm bath}\big\{\gamma _{\mathbf{{q}}^{\prime }}(t^{\prime })\rho ^{\rm bath}\gamma _{%
\mathbf{{q}}^{\prime \prime }}^{\dagger }(t^{\prime \prime
})\big\}\,dt'dt''
\;.
\]
Here we no longer distinguish between the Schr\"{o}dinger and the
interaction picture of the reduced density operator since they
yield the same evolution in the infinite mass limit.
The response function
$\Upsilon (\mathbf{{q}}^{\prime },\mathbf{{q}}^{\prime \prime })$
involves the correlator
\[
\mathrm{tr}_{\rm bath}\big\{\gamma _{\mathbf{{q}}^{\prime }}(t^{\prime })\rho
^{\rm bath}\gamma _{\mathbf{{q}}^{\prime \prime }}^{\dagger }(t^{\prime
\prime })\big\}
=\delta _{\mathbf{{q}}^{\prime }\mathbf{{q}}^{\prime
\prime }}\left| \bar{V}(\mathbf{{q}}^{\prime })\right| ^{2}\sum_{%
\mathbf{{p}}\in\mathbb{P}_\Omega}N_{\mathbf{{p}}}\exp \left( \frac{i(t^{\prime
}-t^{\prime \prime })}{\hbar }\left( \frac{\left| {\mathbf{p}+%
\mathbf{q}^{\prime }}\right| ^{2}-{p}^{2}}{2m}\right) \right)
\]
where
\[
N_{\mathbf{{p}}}=\widetilde{\langle \mathbf{{p}}|} \rho
^{\rm bath}\widetilde{| \mathbf{{p}}\rangle}
\]
is the
probability associated with state $\widetilde{|
\mathbf{{p}}\rangle} $. At this point we include the fact that
$N$ particles exist in volume $\Omega $ by multiplying
the one particle probability by $N$. Moreover,
for a large volume $\Omega$ we can replace the summation over the bath
momenta by an integral,
\[
\sum_{\mathbf{{p}}\in\mathbb{P}_\Omega}
\frac{N_{\mathbf{{p}}}}{\Omega} \,\ldots
\;\rightarrow\;
 \int d\mathbf{p} \,n\mu (\mathbf{p}) \ldots\;,
\]
where $n$ is the bath particle density and $\mu (\mathbf{p})$ the
thermal distribution function (\ref{eq:mu}).
This yields
\begin{eqnarray}
\mathrm{tr}_{\rm bath}\big\{\gamma _{\mathbf{{q}}^{\prime }}(t^{\prime })\rho
^{\rm bath}\gamma _{\mathbf{{q}}^{\prime \prime }}^{\dagger }(t^{\prime
\prime })\big\} &=&\delta _{\mathbf{{q}}^{\prime }\mathbf{{q}}%
^{\prime \prime }} \Omega \left| \bar{V}(\mathbf{{q}}^{\prime })\right|
^{2} \int n \mu (\mathbf{p})\exp \left( \frac{i(t^{\prime }-t^{\prime
\prime })}{\hbar }\left( \frac{\left| \mathbf{p}+\mathbf{{q}}^{\prime
}\right| ^{2}-p^{2}}{2m}\right) \right) d\mathbf{p} \nonumber\\
&\equiv &\delta _{\mathbf{{q}}^{\prime }\mathbf{{q}}^{\prime
\prime }}\Omega G_{\mathbf{{q}}^{\prime }}(t^{\prime \prime
}-t^{\prime }),
\label{eq:trbath}
\end{eqnarray}
with
\begin{equation}
G_{\mathbf{q}}(t^{\prime \prime }-t^{\prime })=n\left| \bar{V}(\mathbf{q}%
)\right| ^{2}\int \mu (\mathbf{p})\exp \left( \frac{i(t^{\prime }-t^{\prime
\prime })}{\hbar }\left( \frac{\left| \mathbf{p}+\mathbf{q}\right| ^{2}-p^{2}%
}{2m}\right) \right) d\mathbf{p}  \label{Fqdef}
\end{equation}
for all $\mathbf{q}$. Then we have
\begin{eqnarray}
\label{eq:deltorhoweak}
\Delta \rho  =
\frac{\left( 2\pi \hbar \right) ^{6}}{\Omega ^{2}}\frac{\Omega }{2\hbar
^{2}}\sum_{\mathbf{{P}}^{\prime },\mathbf{{P}}^{\prime \prime },%
\mathbf{{q}}\in\mathbb{P}_\Omega}\left( 2\tilde{\Lambda}_{\mathbf{{P}}^{\prime }\mathbf{%
{q}}}\rho _{o}\tilde{\Lambda}_{\mathbf{{P}}^{\prime \prime }\mathbf{%
{q}}}^{\dagger }-\tilde{\Lambda}_{\mathbf{{P}}^{\prime \prime }%
\mathbf{{q}}}^{\dagger }\tilde{\Lambda}_{\mathbf{{P}}^{\prime }%
\mathbf{{q}}}\rho _{o}-\rho _{o}\tilde{\Lambda}_{\mathbf{{P}}%
^{\prime \prime }\mathbf{{q}}}^{\dagger }\tilde{\Lambda}_{\mathbf{\tilde{P%
}}^{\prime }\mathbf{{q}}}\right) I_{\mathbf{{q}}}(\Delta t).
\end{eqnarray}
where
\[
I_{\mathbf{q}}(\Delta t)\equiv \int_{0}^{\Delta t}\int_{0}^{\Delta t}G_{%
\mathbf{q}}(t^{\prime \prime }-t^{\prime })\,dt^{\prime }dt^{\prime \prime }
\;.
\]
Now that the Kronecker delta appearing in (\ref{eq:trbath}) has
been summed we can pass to the continuum limit $\Omega\to\infty$
and switch from the box-normalized states (\ref{bigpbar}) to the
standard momentum kets for the Brownian particle,
\begin{eqnarray*}
\left\langle \mathbf{R|P}\right\rangle  &=&\frac{e^{i\mathbf{P\cdot R}/\hbar
}}{\sqrt{(2\pi \hbar )^{3}}}
\;.
\end{eqnarray*}
This is done by the replacements
\[
\frac{\Omega }{(2\pi \hbar )^{3}}\,
\tilde{\Lambda}_{\mathbf{{P}}\mathbf{{q}}}
\;\rightarrow\;
\left| \mathbf{P-q}\right\rangle \left\langle \mathbf{P}\right|
\equiv
\Lambda _{\mathbf{P}\mathbf{q}}
\]
and
\begin{equation}
\frac{(2\pi \hbar )^{3}}{\Omega }
\sum_{\mathbf{{P}}\in\mathbb{P}_\Omega}
\;\rightarrow\;
\int d \mathbf{{P}}
\;.
\label{sumtointegral}
\end{equation}
We obtain
\begin{equation}
\frac{\Delta \rho }{4\pi ^{3}\hbar }=\int d\mathbf{P}^{\prime }d\mathbf{P}%
^{\prime \prime }d\mathbf{q}\left( 2\Lambda _{\mathbf{P}^{\prime }\mathbf{q}%
}\rho _{o}\Lambda _{\mathbf{P}^{\prime \prime }\mathbf{q}}^{\dagger }-\Lambda _{%
\mathbf{P}^{\prime \prime }\mathbf{q}}^{\dagger }\Lambda _{\mathbf{P}^{\prime }%
\mathbf{q}}\rho _{o}-\rho _{o}\Lambda _{\mathbf{P}^{\prime \prime }\mathbf{q}%
}^{\dagger }\Lambda _{\mathbf{P}^{\prime }\mathbf{q}}\right) I_{\mathbf{q}%
}(\Delta t)
\;.  \nonumber
\end{equation}
Similarly,
the Fourier transform of
 (\ref{Vexpansion}) yields in the limit (\ref{sumtointegral})
\begin{eqnarray*}
\bar{V}(\mathbf{q}) &=&\int \frac{d\mathbf{r}}{\left( 2\pi \hbar \right) ^{3}%
}V(\mathbf{r})e^{-i\mathbf{q\cdot r}/\hbar } \\
&=&\left\langle \mathbf{q}|V|\mathbf{0}\right\rangle
\end{eqnarray*}
as the momentum matrix element of the interaction.

By means of the Fourier expansion
% $G_{\mathbf{q}}(t),$%
\begin{equation}
G_{\mathbf{q}}(t)=\int \frac{d\omega }{2\pi }\bar{G}_{\mathbf{q}}(\omega
)e^{-i\omega t}
\label{FqFT}
\end{equation}
we can write
\[
I_{\mathbf{q}}(\Delta t)=\frac{1}{2\pi }\int d\omega \bar{G}_{\mathbf{q}%
}(\omega )\left[ \frac{\sin \left( \omega \Delta t/2\right) }{\omega /2}%
\right] ^{2},
\]
where
\begin{equation}
\bar{G}_{\mathbf{q}}(\omega )=\frac{2\pi nm\hbar }{q}\left| \bar{V}(\mathbf{q%
})\right| ^{2}\left( \frac{\beta }{2\pi m}\right) ^{1/2}\exp \left[ -\frac{%
\beta m\hbar ^{2}}{2q^{2}}\left( \omega -\frac{q^{2}}{2m\hbar }\right)
^{2}\right] ,  \label{Fqomega}
\end{equation}
as we show in Appendix \ref{sec:app2}. It follows that for time
intervals
\begin{equation}
\Delta t\gg \beta \hbar   \label{wcdtcondition}
\end{equation}
we can set $\bar{G}_{\mathbf{q}%
}(\omega )\approx \bar{G}_{\mathbf{q}}(0)$, and $I_{\mathbf{q}}(\Delta
t)\approx \bar{G}_{\mathbf{q}}(0)\Delta t$ for essentially
all $\mathbf{q}$ of importance in (\ref{eq:deltorhoweak}); a coarse graining time long
enough that (\ref{wcdtcondition}) is satisfied is sufficient to guarantee
that Fermi's Golden Rule holds and energy conservation is satisfied for the
scattering bath particles. We then obtain a master equation
for the reduced density operator of the Brownian particle,
\[
\frac{d\rho }{dt}=4\pi ^{3}\hbar \int d\mathbf{P}^{\prime }d\mathbf{P}%
^{\prime \prime }d\mathbf{q\,}\left( 2\Lambda _{\mathbf{P}^{\prime }\mathbf{q}%
}\rho \Lambda _{\mathbf{P}^{\prime \prime }\mathbf{q}}^{\dagger }-\Lambda _{%
\mathbf{P}^{\prime \prime }\mathbf{q}}^{\dagger }\Lambda _{\mathbf{P}^{\prime }%
\mathbf{q}}\rho -\rho \Lambda _{\mathbf{P}^{\prime \prime }\mathbf{q}}^{\dagger
}\Lambda _{\mathbf{P}^{\prime }\mathbf{q}}\right) \bar{G}_{\mathbf{q}}(0).
\]
Although not explicitly in Lindblad form this equation yields a completely
positive evolution of $\rho $ and can be put in Lindblad form, since $\bar{G}%
_{\mathbf{q}}(0)>0$.

 To see the physics of this result and to compare it with earlier work we
go into the coordinate representation,
$\rho (\mathbf{R}_{1},\mathbf{R}_{2};t)=\left\langle \mathbf{R}_{1}|\rho (t)|%
\mathbf{R}_{2}\right\rangle$,
were we find
\[
\frac{d\rho (\mathbf{R}_{1},\mathbf{R}_{2};t)}{dt}=-F(\mathbf{R}_{1}-\mathbf{%
R}_{2})\rho (\mathbf{R}_{1},\mathbf{R}_{2};t),
\]
with
\begin{equation}
F(\mathbf{R})=8\pi ^{3}\hbar \int d\mathbf{q}\left( 1-e^{-i\mathbf{q\cdot R}%
/\hbar }\right) \bar{G}_{\mathbf{q}}(0).  \label{biggamdef}
\end{equation}
While this in practice might be a useful expression for evaluating $F(%
\mathbf{R})$ directly, to compare with earlier work it is easiest to return
first to (\ref{Fqdef}) and formally determine $\bar{G}_{\mathbf{q}}(0)$.
Inserting in (\ref{biggamdef}) leads to
\[
F(\mathbf{R})=16\pi ^{4}\hbar ^{2}n\int d\mathbf{q}\int d
\mathbf{p}\,\mu (\mathbf{p})\left| \bar{V}(\mathbf{q})\right|
^{2}\left[ 1-e^{-i\mathbf{q\cdot R}/\hbar }\right] \delta \left(
\frac{\left| \mathbf{p}+\mathbf{q}\right| ^{2}-p^{2}}{2m}\right) ,
\]
and with the introduction of a new variable $\mathbf{p}^{\prime }=\mathbf{p+q%
}$ to replace $\mathbf{q}$,
\[
F(\mathbf{R})=16\pi ^{4}\hbar ^{2}n\int \int d\mathbf{p}^{\prime }d\mathbf{p}%
\,\mu (\mathbf{p})\left| \bar{V}(\mathbf{p}^{\prime }-\mathbf{p})\right|
^{2}\left[ 1-e^{i(\mathbf{p-p}^{\prime })\mathbf{\cdot R}/\hbar }\right]
\delta \left( \frac{\left( p^{\prime }\right) ^{2}-p^{2}}{2m}\right) .
\]
The $\mathbf{p}^{\prime }$ integration can be done in polar
coordinates, $\mathbf{p}^{\prime }=p^{\prime }\mathbf{\hat{n}}$,
i.e, $d\mathbf{p}^{\prime }=(p^{\prime })^{2}dp^{\prime }d\mathbf{\hat{n
}}$.
Then since
\begin{eqnarray*}
\delta \Big( \frac{\left( p^{\prime }\right) ^{2}-p^{2}}{2m}\Big)
=\delta \Big( \frac{p^{\prime }+p}{2m}(p^{\prime }-p)\Big)
=\frac{m}{p} \,\delta (p^{\prime }-p),
\end{eqnarray*}
we have
\[
F(\mathbf{R})=16\pi ^{4}\hbar ^{2}nm\int \int d\mathbf{\hat{n}}d\mathbf{p}%
\mu (\mathbf{p})p\left| \bar{V}(p\mathbf{\hat{n}}-\mathbf{p})\right|
^{2}\left[ 1-e^{i(\mathbf{p-}p\mathbf{\hat{n}})\mathbf{\cdot R}/\hbar
}\right] ,
\]
which, with the aid of (\ref{muandnu}), yields
\[
F(\mathbf{R)}=4\pi ^{3}\hbar ^{2}mn\int \int p\nu (p)dp\,d\mathbf{\hat{n}\,}d%
\mathbf{\hat{s}}\left| \bar{V}(p\mathbf{\hat{n}}-p\mathbf{\hat{s}})\right|
^{2}\left[ 1-e^{i(p\mathbf{\hat{s}-}p\mathbf{\hat{n}})\mathbf{\cdot R}/\hbar
}\right] .
\]
Finally, we use the fact that the first order Born approximation for the
scattering amplitude is given by \cite{Taylor1972a}
\begin{equation}
f_{\rm B}(\mathbf{p}^{\prime },\mathbf{p)}=-(2\pi )^{2}m\hbar \bar{V}(\mathbf{p}%
^{\prime }-\mathbf{p)}  \label{Born approximation}
\end{equation}
to find
\[
F(\mathbf{R})=n\int \int \frac{p}{m}\nu (p)dp\frac{\,d\mathbf{\hat{n}\,}d%
\mathbf{\hat{s}}}{4\pi }\left| f_{B}(p\mathbf{\hat{n},}p\mathbf{\hat{s})}%
\right| ^{2}\left[ 1-e^{i(p\mathbf{\hat{s}-}p\mathbf{\hat{n}})\mathbf{\cdot R%
}/\hbar }\right] ,
\]
in agreement with (\ref{gamma}) for $\varepsilon =1$, if the full scattering
amplitude is replaced by its first Born approximation.

\section{Conclusions}
\label{sec:conclusions} In this article we presented two detailed
derivations of the quantum master equation for a massive Brownian
particle subject to collisions with particles from a thermalized
environment, and a third argument for the master equation based
on physical motivations. They represent rather different
strategies of dealing with the principal problem that arises in
the formulation of collisional decoherence: The momentum
eigenstates that are the most natural for describing the density
operator of the thermal bath do not constitute proper states for
the application of scattering theory.

By representing the thermal bath with an appropriate over-complete basis, we
obtained in section \ref{sec:scatt} the full master equation in a calculation that is
mathematically and physically convincing, but cumbersome. The perturbative
treatment in section ~\ref{sec:weak} permitted the avoidance of improper states up to a
point where they could be handled in an unambiguous way. Yet this
calculation hides some of the physics involved, and does not in itself
suggest an immediate generalization to higher orders in the perturbation
parameter. More straightforward, but also most ventured, is the use of the
replacement rule put forward in section ~\ref{sec:dsquared}. It yields the master equation
immediately, but lacks the mathematical foundation one would wish for an
equation describing a process as fundamental as decoherence by collisions.
The strategies developed here lead to natural generalizations, both to the
more general problem of a Brownian particle of finite mass, and to other
collisional decoherence processes. We plan to turn to them in future
communications.

Different as the three approaches may be, they all suggest that previous
results in the literature predict decoherence rates that are quantitatively
too large. This conclusion is not only of theoretical interest. A recent
experiment \cite{Hornberger2003a} on the decoherence of fullerene matter waves by collisions
with background gas atoms was sensitive to the presence or absence of the
factor of $2\pi $ that has been the focus of this article. And indeed, the
observed decoherence rates are in full agreement with the equation derived
in this article, and exclude the previous results.

\begin{acknowledgments}

JS acknowledges support from the Natural Sciences and Engineering Research
Council of Canada, and thanks the University of Vienna for its hospitality
during his stay as Visiting Professor. KH has been supported
by the DFG Emmy Noether program.
\end{acknowledgments}

\appendix
\section{}
\label{sec:app1}

In this Appendix we confirm the results (\ref{I1result},\ref{I2result}),
beginning from the definitions (\ref{Idef}). We begin by inserting a
complete set of momentum eigenstates to write the matrix element appearing
in $I_{2}(\mathbf{R})$ as
\begin{eqnarray*}
\left\langle \mathbf{q}_{2}|\mathcal{T}_{o}^{\dagger }e^{i\mathfrak{p}\cdot
\mathbf{R}/\hbar }\mathcal{T}_{o}|\mathbf{q}_{1}\right\rangle  &=&\int d%
\mathbf{q}^{\prime }\left\langle \mathbf{q}_{2}|\mathcal{T}_{o}^{\dagger
}e^{i\mathfrak{p}\cdot \mathbf{R}/\hbar }|\mathbf{q}^{\prime }\right\rangle
\left\langle \mathbf{q}^{\prime }|\mathcal{T}_{o}|\mathbf{q}%
_{1}\right\rangle  \\
&=&\int d\mathbf{q}^{\prime }e^{i\mathbf{q}^{\prime }\cdot \mathbf{R}/\hbar
}\left\langle \mathbf{q}_{2}|\mathcal{T}_{o}^{\dagger }|\mathbf{q}^{\prime
}\right\rangle \left\langle \mathbf{q}^{\prime }|\mathcal{T}_{o}|\mathbf{q}%
_{1}\right\rangle .
\end{eqnarray*}
Then using the expression (\ref{Tnought}) for the matrix elements of $%
\mathcal{T}_{o}$ and, noting that we can write $\delta (q_{2}-q^{\prime
})\delta (q^{\prime }-q_{1})=\delta (q_{2}-q_{1})\delta (q^{\prime }-q_{1})$%
, we find
\begin{eqnarray}
I_{1} &=&\int \frac{d\mathbf{q}_{1}d\mathbf{q}_{2}}{2\pi \hbar q_{2}}u(%
\mathbf{q}_{1},\mathbf{q}_{2})\left( f(\mathbf{q}_{2},\mathbf{q}_{1})+f^{*}(%
\mathbf{q}_{1},\mathbf{q}_{2})\right) \delta (q_{2}-q_{1}),  \label{Itemps}
\\
I_{2}(\mathbf{R}) &=&\int \frac{d\mathbf{q}_{1}d\mathbf{q}_{2}d\mathbf{q}%
^{\prime }}{4\pi ^{2}\hbar ^{2}q_{1}q_{2}}u(\mathbf{q}_{1},\mathbf{q}%
_{2})e^{i\mathbf{q}^{\prime }\cdot \mathbf{R}/\hbar }f^{*}(\mathbf{q}%
^{\prime },\mathbf{q}_{2})f(\mathbf{q}^{\prime },\mathbf{q}_{1})\delta
(q^{\prime }-q_{1})\delta (q_{2}-q_{1}),  \nonumber
\end{eqnarray}
and so $\delta (q_{2}-q_{1})$ appears in both these expressions.
They are simplified if we make a change of variables from
$(\mathbf{q}_{1}, \mathbf{q}_{2})$ to $(\mathbf{q},\mathbf{s})$,
where $\mathbf{q}_{2} =\mathbf{q+}{\mathbf{s}}/{2}$ and
$\mathbf{q}_{1} =\mathbf{q-}{\mathbf{s}}/{2}$.
The Jacobian of this transformation is unity, so $d\mathbf{q}_{1}d\mathbf{q}%
_{2}=d\mathbf{q}d\mathbf{s}$, and
\[
\delta (q_{2}-q_{1})=\delta \left( \sqrt{q^{2}+\frac{s^{2}}{4}+\mathbf{%
s\cdot q}}-\sqrt{q^{2}+\frac{s^{2}}{4}-\mathbf{s\cdot q}}\right) .
\]
We do the $\mathbf{s}$ integral first, and we single out the component of $%
\mathbf{s}$ along $\mathbf{\hat{q}}$ by writing
\[
\mathbf{s=}\nu \mathbf{\hat{q}+\Delta },
\]
where $\mathbf{\Delta }$ is a two-dimensional vector lying in the plane
perpendicular to $\mathbf{\hat{q}}$. Then $d\mathbf{s}=d\nu d\mathbf{\Delta }
$, and
\[
\delta (q_{2}-q_{1})=\delta (g(\nu ))
\]
for fixed $\mathbf{\Delta }$, where
\[
g(\nu )=\sqrt{q^{2}+\frac{\Delta ^{2}+\nu ^{2}}{4}+q\nu }-\sqrt{q^{2}+\frac{%
\Delta ^{2}+\nu ^{2}}{4}-q\nu }
\]
which as a function of $\nu $ vanishes at $\nu =0$. Thus
\[
\delta (g(\nu ))=\frac{\delta (\nu )}{\left| dg/d\nu \right| _{\nu =0}}=%
\frac{\sqrt{q^{2}+\frac{\Delta ^{2}}{4}}}{q}\delta (\nu )
\;,
\]
and for any function $U(\mathbf{q}_{1},\mathbf{q}_{2})$ we have
\begin{eqnarray}
&&d\mathbf{q}_{1}d\mathbf{q}_{2}U(\mathbf{q}_{1},\mathbf{q}_{2})\delta
(q_{2}-q_{1})  \label{red} \\
&=&d\mathbf{q}d\nu d\mathbf{\Delta \,}U(\mathbf{q}-\frac{\mathbf{s}}{2},%
\mathbf{q}+\frac{\mathbf{s}}{2})\delta (g(\nu ))  \nonumber \\
&=&d\mathbf{q}d\mathbf{\Delta }\frac{Q}{q}\mathcal{\,}U(\mathbf{q}-\frac{%
\mathbf{\Delta }}{2},\mathbf{q}+\frac{\mathbf{\Delta }}{2})
\;,
\nonumber
\end{eqnarray}
where we have put
\[
Q\equiv \sqrt{q^{2}+\frac{\Delta ^{2}}{4}}=\left| \mathbf{q-}\frac{\mathbf{%
\Delta }}{2}\right| =\left| \mathbf{q+}\frac{\mathbf{\Delta }}{2}\right|
=q_{1}=q_{2}
\;,
\]
\textit{cf. }equation (\ref{Qdef}). Using this in the first of
(\ref{Itemps}) we immediately find (\ref{I1result}); using it in
the second of (\ref{Itemps}) we find  (\ref{I2result})
\begin{eqnarray*}
I_{2}(\mathbf{R)} &=&\int \frac{d\mathbf{q}d\mathbf{q}^{\prime }}{4\pi
^{2}\hbar ^{2}}\int_{\mathbf{\hat{q}}^{\perp }}\frac{d\mathbf{\Delta }}{qQ}u(%
\mathbf{q-}\frac{\mathbf{\Delta }}{2},\mathbf{q+}\frac{\mathbf{\Delta }}{2}%
)e^{i\mathbf{q}^{\prime }\cdot \mathbf{R}/\hbar }f^{*}(\mathbf{q}^{\prime },%
\mathbf{q+}\frac{\mathbf{\Delta }}{2})f(\mathbf{q}^{\prime },\mathbf{q-}%
\frac{\mathbf{\Delta }}{2})\delta (q^{\prime }-Q) \\
&=&\int \frac{d\mathbf{q}d\mathbf{\hat{n}}}{4\pi ^{2}\hbar ^{2}}\int_{%
\mathbf{\hat{q}}^{\perp }}d\mathbf{\Delta }\frac{Q}{q}u(\mathbf{q-}\frac{%
\mathbf{\Delta }}{2},\mathbf{q+}\frac{\mathbf{\Delta }}{2})e^{i\mathbf{Q}%
\cdot \mathbf{R}/\hbar }f^{*}(\mathbf{Q},\mathbf{q+}\frac{\mathbf{\Delta }}{2%
})f(\mathbf{Q},\mathbf{q-}\frac{\mathbf{\Delta }}{2}),
\end{eqnarray*}
where to get from the first to the second line we have put $d\mathbf{q}%
^{\prime }=(q^{\prime })^{2}dq^{\prime }d\mathbf{\hat{n}}$, where $d\mathbf{%
\hat{n}}$ is an element of solid angle, and defined $\mathbf{Q}$ as in (\ref{Qdef}).

\section{}
\label{sec:app2}

Here we confirm the result (\ref{Fqomega}). From the inverse transform of (%
\ref{FqFT}) we have, using (\ref{Fqdef}),
\begin{eqnarray}
\bar{G}_{\mathbf{q}}(\omega ) &\equiv &\int dt\,e^{i\omega t}G_{\mathbf{q}%
}(t)  \label{Fqomegatemp} \\
&=&2\pi \hbar n\left| \bar{V}(\mathbf{q})\right| ^{2}\int \mu (\mathbf{p}%
)\delta \left( \frac{\left| \mathbf{p}+\mathbf{q}\right| ^{2}-p^{2}}{2m}%
-\hbar \omega \right) d\mathbf{p}  \nonumber \\
&=&\pi \hbar \left| \bar{V}(\mathbf{q})\right| ^{2}n\int \nu
(p)\,dp\,d\alpha \,\delta \left( s_{\omega }(\alpha )\right)   \nonumber
\end{eqnarray}
where
\[
s_{\omega }(\alpha )=\frac{2pq\alpha +q^{2}}{2m}-\hbar \omega .
\]
To get from the second to the third line of (\ref{Fqomegatemp}) we have put
\begin{eqnarray*}
\frac{\left| \mathbf{p}+\mathbf{q}\right| ^{2}-p^{2}}{2m} &=&\frac{2\mathbf{%
p\cdot q}+q^{2}}{2m} \\
&=&\frac{2pq\alpha +q^{2}}{2m},
\end{eqnarray*}
where $\alpha =\cos \theta $, and $\theta $ is the angle between $\mathbf{p}$
and $\mathbf{q}$, and used (\ref{muandnu}), writing $d\mathbf{\hat{s}}%
=d\alpha d\phi $, where $\phi $ is the azimuthal angle around $\mathbf{q}$.
The function $s_{\omega }(\alpha )$ has a single root
\[
\alpha _{o}=\frac{2m\hbar \omega -q^{2}}{2pq}
\;,
\]
which, for there to be a contribution to the integral in (\ref{Fqomegatemp}),
must satisfy $-1\leq \alpha
_{o}\leq 1$.
Since $p$ and $q$  are both positive this implies
\[
p\geq p_\mathrm{cut}\equiv \frac{\left| 2m\hbar \omega -q^{2}\right| }{2q}
\;.
\]
Finally, noting that
\[
\delta (s_{\omega }(\alpha ))=\frac{m\delta (\alpha -\alpha _{o})}{pq}
\;,
\]
we obtain the expression
\begin{equation}
\bar{G}_{\mathbf{q}}(\omega )=\frac{\pi m\hbar }{q}\left| \bar{V}(\mathbf{q}%
)\right| ^{2}n\int_{p_\mathrm{cut}}^{\infty }\frac{\nu
(p)}{p}\,dp\;.
\nonumber
\end{equation}
In thermal equilibrium,
\[
\nu (p)=4\pi \left( \frac{\beta }{2\pi m}\right) ^{3/2}e^{-\beta
p^{2}/(2m)}p^{2},
\]
this integral
immediately yields (\ref{Fqomega}).

%\bibliographystyle{apsrev}
%\bibliography{mqo}
%\bibitem{WeissChap2}
%  see, \emph{e.g.,} chapter 2 of
%\bibinfo{author}{\bibfnamefont{U.} \bibnamefont{Weiss}},
%  \emph{\bibinfo{title}{Quantum Dissipative Systems}}, 2nd edition,
%  (\bibinfo{publisher}{World Scientific},
%  \bibinfo{address}{Singapore}, \bibinfo{year}{1999}), and
%  references therein.
%\bibitem{Kottos2003a}
%\bibinfo{author}{\bibfnamefont{T.}~\bibnamefont{Kottos}} \bibnamefont{and}
%  \bibinfo{author}{\bibfnamefont{U.} \bibnamefont{Smilansky}},
%  \bibinfo{journal}{J. Phys. A} (\bibinfo{year}{2003}),
%  to be published, \\ preprint:  arXiv:nlin.CD/0207049.

\end{document}